\documentclass[12pt]{article}
\usepackage{makeidx}
\usepackage{times}
\usepackage{amssymb}
\usepackage{amsmath}
\numberwithin{equation}{section}
\usepackage{setspace}
\usepackage{hyperref}
\usepackage{color}
\usepackage{graphicx}
\usepackage{rotating}
\usepackage{pdflscape}
\usepackage{epstopdf}
\usepackage[round,authoryear,comma]{natbib}
\usepackage{natbib,hyperref}
\usepackage{booktabs}
\usepackage{setspace}
\usepackage{pgfplots}
\usepackage{subcaption}
\usepackage{xcolor}
\usepackage{multirow}
\usepackage{array}
\usepackage{placeins}
\usepackage{actuarialangle}
\hypersetup{
	colorlinks   = true, 
	urlcolor     = blue, 
	linkcolor    = blue, 
	citecolor   = blue 
}

\newlength{\Oldarrayrulewidth}

\providecommand{\U}[1]{\protect\rule{.1in}{.1in}}
\newtheorem{theorem}{Theorem}[section]

\newtheorem{corollary}{Corollary}

\newtheorem{definition}{Definition}[section]

\newtheorem{lemma}{Lemma}[section]

\parskip=\medskipamount

\newcolumntype{C}[1]{>{\centering\arraybackslash}m{#1}}

\begin{document}

\title{Dependence bounds for the difference of stop-loss payoffs on the difference of two random variables}
\date{ }
\author{Hamza Hanbali\thanks{Monash University, email address: \texttt{hamza.hanbali@monash.edu}}\and Jan Dhaene\thanks{KU Leuven, email address: \texttt{jan.dhaene@kuleuven.be}}  \and
	 	Dani\"{e}l Linders\thanks{University of Amsterdam, email address: \texttt{d.h.linders@uva.nl}}}
\maketitle
\begin{center}\rule{10cm}{0.5pt}\end{center}
\begin{abstract}
	\noindent
This paper considers the difference of stop-loss payoffs where the underlying is a difference of two random variables. The goal is to study whether the comonotonic and countermonotonic modifications of those two random variables can be used to construct upper and lower bounds for the expected payoff, despite the fact that the payoff function is neither convex nor concave. The answer to the central question of the paper requires studying the crossing points of the cumulative distribution function (cdf) of the original difference with the cdf's of its comonotonic and countermonotonic transforms. The analysis is supplemented with a numerical study of longevity trend bonds, using different mortality models and population data. The numerical study reveals that for these mortality-linked securities the three pairs of cdf's generally have unique pairwise crossing points. Under symmetric copulas, all crossing points can reasonably be approximated by the difference of the marginal medians, but this approximation is not necessarily valid for asymmetric copulas. Nevertheless, extreme dependence structures can give rise to bounds if the layers of the bond are selected to hedge tail risk. Further, the dependence uncertainty spread can be low if the layers are selected to hedge median risk, and, subject to a trade-off, to hedge tail risk as well.


\end{abstract}
\newpage
\section{Introduction}
\label{Introduction}
The expectation of hedging instruments with a convex or concave payoff function of multiple risks, commonly the expected stop-loss payoff of sums or differences of random variables, attains its dependence bounds under extreme dependence structures of the underlying risks. For the difference of two random variables with arbitrary dependence structure, where the difference could represent the hedger's net profit or a spread between two indices, the upper bound of the expected stop-loss payoff is attained in the countermonotonic case whereas the lower bound is attained in the comonotonic case \citep{Dhaene2002a,Dhaene2002b,Chaoubi}. These inequalities are linked to stochastic orders and have meaningful economic interpretations in that risk-averse decision-makers under the expected utility framework prefer diversification over concentration \citep{DenuitDhaeneGoovaertsKaas2005,ShakedShanthihumar}. For hedging instruments with convex or concave payoff function, expected payoffs under extreme dependence structures can be used as dependence model-free bounds in the pricing process \citep{MeileisonNadas1979,Dhaene2000,Hobson2005}, or to evaluate the extent of dependence model risk using the dependence uncertainty spread which is the difference between the upper and lower bounds \citep{Bignozzi_et_al_2015,Chaoubi}. 

An important hedging strategy corresponds to the difference of two expected stop-loss payoffs with two different retentions. This strategy is known as \textit{bull call spread} in the quantitative finance literature. In the actuarial literature, it is known as \textit{layer reinsurance}, or \textit{limited stop-loss} \citep{Wang1996,ChiTan,CaiWeng}. This payoff structure also appears in many alternative risk transfers, such as mortality-linked securities, where the underlying is a linear combination of some relevant indices. For a given random variable $I$, the payoff of this hedging strategy with underlying $I$ is denoted by $r(I)$, such that the payoff function $r$ is given by:
\begin{equation}\label{Eq1}
	r:x\mapsto \frac{B}{\epsilon-\delta}\left( \left(x-\delta\right)_+ - \left(x-\epsilon\right)_+ \right),
\end{equation}
where $\delta<\epsilon$, and $B$ is the maximum payable amount. This payoff function is neither globally convex, nor globally concave. Thus, when it involves multiple risks, it is unclear whether extreme dependence structures lead to upper and lower bounds, and whether they can be used for the management and measurement of dependence model risk. 

The present paper addresses this problem for the case where the underlying $I$ corresponds to the difference of two random variables $I_1$ and $I_2$, i.e. $I=I_1-I_2$. The aim of the paper is to investigate whether comonotonicity and countermonotonicity of $(I_1,I_2)$ lead to bounds for the expected value of the payoff $r(I)$. The problem of interest to the present paper can be formulated in the framework of stochastic ordering. Namely, denoting by $I^c$ and $I^{cm}$ the comonotonic and countermonotonic modifications of $I$, respectively, it is well-known that $I^c\preceq_{cx} I \preceq_{cx} I^{cm}$ for any dependence structure of $\left(I_1,I_2\right)$, where `$\preceq_{cx}$' denotes the convex order \citep{Ruschendorf1983,Dhaene_Denuit_1999,Kaas_etal:Upper_Lower_Bounds}. In other words, despite all having the same expected value, $I^{cm}$ is riskier than $I$, and $I$ is riskier than $I^c$. Further, risk-averse decision-makers under classical expected utility theory prefer the risk $I^c$ over the risk $I$, and the risk $I$ over the risk $I^{cm}$. The present paper studies whether the transformation $r$ leads to an order of the expected payoffs which is consistent with $I^c\preceq_{cx} I \preceq_{cx} I^{cm}$. In particular, the central question is whether $r$ preserves or reverses these inequalities in terms of expectations.

The contribution of this paper is threefold. First, it studies conditions under which the transformation $r$ given by \eqref{Eq1} preserves or reverses the order of expected payoffs for a difference of random variables. Second, the paper studies the crossing points of the cdf's $F_{I}$, $F_{I^c}$ and $F_{I^{cm}}$, which turn out to be essential in the analysis, and derives analytical properties. Third, the paper investigates the problem in the context of mortality-linked securities with a focus on longevity-trend bonds by conducting a large numerical analysis using six mortality models and population data from 21 countries.

Upper and lower bounds of various risk measures have been derived in a substantial number of studies whose common aim is addressing dependence model risk. Many contributions in this direction have used extreme dependence structures to derive model-free bounds for convex or concave functions of multiple risks \citep{Dhaene_Denuit_1999,Dhaene2000,Dhaene2002a,Dhaene2002b,Hobson2005,CheungLo2013,Cheung2017,Chaoubi}. A related stream of the literature addresses the problem by incorporating additional information or by using numerical algorithms, which can also be useful for some non-convex/non-concave functions \citep{Kaas_etal:Upper_Lower_Bounds,Embrechts2005,Mesfioui,WangPengYang2013,Villegas2012,Puccetti2013,BernardJRI}. When extreme dependence structures lead to sharp bounds, an advantage of using them is that the expected payoff can be decomposed in terms of the expected payoffs of the individual components. In this case, the value of the bound can be determined from prices quoted in the market. This approach provides model-free safeguards for pricing more exotic instruments using vanilla ones when the market is mature and sufficiently liquid. Decomposition formulas for expected stop-loss payoffs of comonotonic sums have been studied in \cite{MeileisonNadas1979}, \cite{Dhaene2000}, \cite{Hobson2005} and \cite{Linders2012}, among others, and were applied recently in \cite{Bahl2016} for catastrophic mortality bonds. Decomposition formulas for expected stop-loss payoffs of countermonotonic sums were studied in \cite{LaurenceWang2009} and \cite{HanbaliLindersDhaene}. To the best of knowledge, no previous studies have considered the question of whether extreme dependence structures lead to bounds for the payoff function $r$.

The present paper adds to this literature by investigating when comonotonicity and countermonotonicity can be used as bounds for the payoff function \eqref{Eq1}, and when they cannot. The analysis shows that the choice of $\delta$ and $\epsilon$ is crucial for extreme dependence structures to be appropriate bounds. Depending on the values of these parameters, comonotonicity and countermonotonicity may lead to two upper bounds, or to two lower bounds, which would render the analysis of dependence uncertainty spread using extreme dependence structures meaningless for such payoff functions. When these values are such that extreme dependence structures are two opposite bounds, the choice of $\delta$ and $\epsilon$ also determines which of comonotonicity and countermonotonicity is the upper or lower bound. The relevance of extreme dependence structures as model-free dependence bounds may also be hindered by the fact that the bounds can be \textit{dependence-dependent}. In particular, comonotonicity and countermonotonicity may be appropriate for some dependence structures of $\left(I_1,I_2\right)$, but not for all. 

The intervals of $\delta$ and $\epsilon$ that unveil the order of expected payoffs are defined by the crossing points of the cdf's $F_{I}$, $F_{I^c}$ and $F_{I^{cm}}$. There exists an extensive literature on crossing points of cdf's. This literature revolves around notions of mean-preserving spread and stochastic dominance; see e.g.\ \cite{RothschildStiglitz1970}, \cite{Cohen1995}, \cite{Chateauneuf2004} and \cite{Muller2017}. Statistical tests to ascertain the existence of crossing points can be found in \cite{HawkinsKochar1991}, \cite{ChenChenChen2002}, and \cite{AdamFerger2012}, among others. These tests are tailored to identify stochastic dominance orders, but they are not relevant in the present context, because it is already known that the cdf's $F_I$, $F_{I^c}$ and $F_{I^{cm}}$ cross at least once. Specifically, the problem of concern here is not the existence of the crossing points, which is guaranteed from the convex order properties, but instead, it is the number of crossing points and their analytical expressions, which is a complex issue to investigate in general.

An ideal situation that simplifies the problem is when the cdf's of $I$, $I^c$ and $I^{cm}$ have unique pairwise crossing points which coincide for any joint distribution of $\left(I_1,I_2\right)$. The present paper reports a numerical study using a large palette of continuous marginal distributions and copula models, which shows that in many cases, the pairwise crossing points are unique, but they do not necessarily coincide. However, even though this situation is closer to the ideal one, it does not always hold.

The present paper adds to the literature on crossing points of cdf's by deriving some properties of the crossing points of $F_I$, $F_{I^c}$ and $F_{I^{cm}}$. Namely, in case the three cdf's have unique pairwise crossing points, an order of these points is established. In case the indices $I_1$ and $I_2$ are ordered in the dispersive order sense, it is proved that the crossing point of the comonotonic and the countermonotonic cdf's is unique, and is equal to the difference of the marginal medians. In case the indices $I_1$ and $I_2$ satisfy appropriate conditions of symmetry, it is proved that all three cdf's have a unique common crossing point which is again equal to the difference of the marginal medians. This latter result implies that, under the conditions of symmetry, choosing $\delta$ and $\epsilon$ above the median means that comonotonicity leads to a lower bound and countermonotonicity leads to an upper bound, and the opposite holds if $\delta$ and $\epsilon$ are both below the median.

Finally, the paper investigates the crossing points of $F_I$, $F_{I^c}$ and $F_{I^{cm}}$ in the context of mortality-linked securities, with a focus on longevity trend bonds, where the construction of the indices $I_1$ and $I_2$ is similar to that of Swiss Re's 2010 Kortis bond \citep{HuntBlake2015,LiTang2019}. A large numerical study is conducted using mortality data from 21 countries, and combinations of several copula models with six different mortality models. The analysis reveals that the pairwise crossing points are unique under all models considered in this paper. For symmetric copulas, the pairwise crossing points can be approximated by the difference of the marginal medians. For asymmetric copulas, such as the Clayton copula, this approximation does not hold in general. Nevertheless, the results suggest that if the values of $\delta$ and $\epsilon$ are chosen to hedge against tail risk, i.e. for sufficiently large or sufficiently small $\delta$ and $\epsilon$, then the countermonotonic and comonotonic transforms can lead to upper and lower bounds. Further, when $\delta$ and $\epsilon$ are chosen to hedge against median risk, and to some extent tail risk as well, the analysis of the expected payoffs suggests that the dependence uncertainty spread can be close to zero, meaning that dependence model risk can be low. However, for tail risk, the latter conclusion is subject to a trade-off: higher layers allow to reduce the dependence uncertainty spread, but if they are too high, the longevity trend bond may not be relevant anymore because the layer it intends to hedge would be too large for the support of $I$.

The remainder of this paper is organized a follows. Section \ref{Section-2} introduces the main concepts and notations, and provides details on the main problem addressed in the paper. Section \ref{Sec-Bounds} derives conditions under which comonotonicity and countermonotonicity constitute bounds under the transformation $r$. Section \ref{Sec-Crossing} is devoted to the crossing points of the cdf's $F_I$, $F_{I^c}$ and $F_{I^{cm}}$. Section \ref{Sec-NumericalAnalysis} reports the numerical analysis for longevity trend bonds. Section \ref{Sec-Conclusion} concludes the paper. The appendix contains the proofs and details on the mortality models.

\section{Preliminaries}
\label{Section-2}
This section lays out the concepts and properties which are relevant in this paper; more details can be found in e.g. \cite{DenuitDhaeneGoovaertsKaas2005} and \cite{ShakedShanthihumar}. Subsection \ref{Sec21} introduces convex order. Subsection \ref{Sec22} introduces comonotonicity and countermonotonicity. Subsection \ref{Sec23} presents the problem of interest.

Throughout the paper, all random variables (henceforth r.v.'s) are defined on a common probability space and are absolutely continuous with strictly increasing cumulative distribution function (henceforth cdf), and with finite expectations. The cdf of the real-valued r.v.\ $X$ is denoted by $F_X$, and its inverse cdf is denoted by $F_X^{-1}$. The infimum and supremum of $X$ are denoted by $l_X=F^{-1}_X(0)$ and $u_X=F^{-1}_X(1)$, respectively. The expectation $\mathbb{E}\left[\left(X-x\right)_+\right]$ is referred to as the upper tail transform of $X$ at the level $x$, or expected stop-loss payoff of $X$ with retention $x$.

\subsection{Convex order}\label{Sec21}

\begin{definition}[Convex order]\label{Def:Convex}
	The r.v.\ $Y$ is said to be larger than $X$ in the convex order, with the notation $X\preceq_{cx} Y$, if and only if $\mathbb{E}\left[  u(-Y)\right]  \leq \mathbb{E}\left[  u(-X)\right]$ for all concave functions $u$, provided the expectations exist, or equivalently, if and only if $\mathbb{E}\left[  X\right]  =\mathbb{E}\left[  Y\right]$ and
	\begin{equation}\label{ec38}
		\mathbb{E}\left[  (X-x)_{+}\right]  \ \leq\mathbb{E}\left[  (Y-x)_{+}\right]
		\text{,\quad for all }x\in\mathbb{R}.
	\end{equation}
\end{definition}

If $X \preceq_{cx} Y$, the cdf's of $X$ and $Y$ cross at least once. The number of crossing points must be odd in case it is finite, with $F_{X} \leq F_{Y}$ before the first crossing point, $F_{X} \geq F_{Y}$ after the last crossing point, and the sign of $F_X-F_Y$ alternates after each crossing point; see Theorem 3.A.5 of \cite{ShakedShanthihumar}, as well as \cite{Shaked1980} and \cite{Hesselager}. In some situations, the cdf's $F_X$ and $F_Y$ may be equal over an entire interval, and the definition of the crossing points is too broad. Further, the cdf's may be equal over an interval but the crossing would not occur. The set $D(X,Y)$ defined hereafter specifies the crossing points which are relevant in the present context. Note that, implicitly, it is assumed that the cdf's $F_X$ and $F_Y$ are distinct. In the trivial case where $X\overset{d}{=}Y$, the set $D(X,Y)$ is empty, by convention.
\begin{definition}[crossing points]
	Let $D(X,Y)$ be the set of $N$ crossing points $d_1<d_2<...<d_N$ of the cdf's $F_X$ and $F_Y$, with $l_X<d_1$ and $d_N<u_X$, where $l_X=F^{-1}_X(0)$ and $u_X=F^{-1}_X(1)$.
	\begin{itemize}
		\item For $i = 2m$ where $m\in \mathbb{N}$, $d_i\in D(X,Y)$ if and only if there exist $a_i,b_i,c_i\in (l_X,u_X)$ with $a_i<b_i\leq d_i<c_i$ such that $F_{Y}(x)<F_{X}(x)$ for $x\in[a_i,b_i)$, and $F_{Y}(x)=F_{X}(x)$ for $x\in[b_i,d_i]$, and $F_{Y}(x)>F_{X}(x)$ for $x\in (d_i,c_i]$.
		\item For $i = 2m+1$ where $m\in \mathbb{N}$, $d_i\in D(X,Y)$ if and only if there exist $a_i,b_i,c_i\in (l_X,u_X)$ with $a_i<b_i\leq d_i<c_i$ such that $F_{Y}(x)>F_{X}(x)$ for $x\in[a_i,b_i)$, and $F_{Y}(x)=F_{X}(x)$ for $x\in[b_i,d_i]$, and $F_{Y}(x)<F_{X}(x)$ for $x\in (d_i,c_i]$.
	\end{itemize}
\end{definition}
In particular, on top of the condition $F_{X}(d_i)=F_{Y}(d_i)$, the above definition requires strict inequalities on some sub-intervals that alternate from $[d_{i-1},d_i]$ to $[d_i,d_{i+1}]$.

The sets $D^{(j)}_{\leq}(X,Y)$, for $j=0,1,\ldots,\frac{N-1}{2}$, include all $x\in (l_X,u_X)$ such that $F_X(x)\leq F_Y(x)$:
\begin{equation}\label{Set3}D^{(0)}_{\leq} = (l_X,d_1] \qquad \text{and } \qquad D^{(j)}_{\leq}(X,Y) = [d_{2j},d_{2j+1}], \quad \text{for } j=1,\ldots,(N-1)/2,\end{equation}
whereas $D^{(j)}_{\geq}(X,Y)$, for $j=0,1,\ldots,\frac{N-1}{2}$, include all $x\in (l_X,u_X)$ such that $F_X(x)\geq F_Y(x)$:
\begin{equation}\label{Set4}D^{(0)}_{\geq} = [d_N,u_X)\qquad \text{and} \qquad D^{(j)}_{\geq}(X,Y) = [d_{2j-1},d_{2j}], \quad \text{for } j=1,\ldots,(N-1)/2.\end{equation}
By convention, when $N=1$, $D^{(j)}_{\leq}(X,Y)=D^{(j)}_{\geq}(X,Y)=\emptyset$ for $j\geq1 $, and when $X\overset{d}{=}Y$, $D^{(0)}_{\leq}(X,Y)=D^{(0)}_{\geq}(X,Y)=(l_X,u_X)$ and $D^{(j)}_{\leq}(X,Y)=D^{(j)}_{\geq}(X,Y)=\emptyset$ for $j\geq 1$.

\subsection{Comonotonicity and countermonotonicity}\label{Sec22}
\begin{definition}[Comonotonic and countermonotonic modifications]
	For $U$ a uniform r.v.\ on $[0,1]$, the comonotonic and countermonotonic modifications of $\left(  I_{1},I_{2}\right)$ are denoted by $\left(  I_{1}^{c},I_{2}^{c}  \right)$ and $\left(  I_{1}^{cm},I_{2}^{cm}  \right)$, and represented as follows:
$$
		\left(  I_{1}^{c},I_{2}^{c}  \right)\overset{d}{=}\left(  F_{I_{1}}^{-1}\left(  U\right)
		,F_{I_{2}}^{-1}\left(U\right)  \right) \quad \text{and} \quad \left(  I_{1}^{cm},I_{2}^{cm}  \right)\overset{d}{=}\left(  F_{I_{1}}^{-1}\left(  U\right)
		,F_{I_{2}}^{-1}\left(1-  U\right)  \right).%
$$
\end{definition}
For a random vector $(I_1,I_2)$ with fixed marginal cdf's $F_{I_1}$ and $F_{I_2}$ and joint cdf $F_{\left(I_1,I_2\right)}$, the Fr\'{e}chet space $\mathcal{F}\left(F_{I_1},F_{I_2}\right)$ is the set of all possible joint cdf's $F_{\left(I_1,I_2\right)}$. The bivariate cdf's resulting from the comonotonic and countermonotonic transformations of $(I_1,I_2)$ belong to the Fr\'{e}chet space $\mathcal{F}\left(F_{I_1},F_{I_2}\right)$. In particular, these modifications are such that the marginal distributions $F_{I_1}$ and $F_{I_2}$ remain unchanged, but the dependence structure is transformed into an extreme positive (comonotonic) or negative (countermonotonic) dependence structure. 

\subsection{Main problem}\label{Sec23}
The quantity of interest in the present paper is a difference of r.v.'s, namely the r.v.\ $I$ with $I=I_1-I_2$. The notation $\mathcal{I}\left(F_{I_1},F_{I_2}\right)$ is used to denote the set of all univariate cdf's $F_I$ of the random difference $I$ obtained from all possible joint cdf's in the Fr\'{e}chet space $\mathcal{F}\left(F_1,F_2\right)$. Let $I^c$ and $I^{cm}$ be the differences obtained from the modifications $\left(I_1^c,I_2^c\right)$ and $\left(I_1^{cm},I_2^{cm}\right)$, such that:
$$	I^{c}\overset{d}{=}F_{I_1}^{-1}(U)-F_{I_2}^{-1}(U),\qquad \text{and} \qquad I^{cm}\overset{d}{=}F_{I_1}^{-1}(U)-F_{I_2}^{-1}(1-U).$$
Naturally, the cdf's $F_{I^c}$ and $F_{I^{cm}}$ belong to $\mathcal{I}\left(F_{I_1},F_{I_2}\right)$. Further, the following inequalities hold:
\begin{equation}\label{Eq3-15}
I^c \preceq_{cx} I \preceq_{cx} I^{cm},
\end{equation}
and they do so for any random difference $I$ with cdf $F_I \in \mathcal{I}\left(F_{I_1},F_{I_2}\right)$. Relevant references for these inequalities include \cite{Ruschendorf1983}, \cite{Dhaene_Denuit_1999}, \cite{Dhaene2000}, \cite{Kaas_etal:Upper_Lower_Bounds}, and \cite{Chaoubi}. Taking into account Inequality \eqref{Eq3-15} as well as Definition \ref{Def:Convex} of convex order, and noting that the stop-loss payoff is convex, the following holds:
\begin{equation}\label{ConvexIneq}
	\mathbb{E}\left[\left(I^c-x\right)_+\right]\leq\mathbb{E}\left[\left(I-x\right)_+\right]\leq \mathbb{E}\left[\left(I^{cm}-x\right)_+\right], \qquad \text{for all } x\in\mathbb{R}.
\end{equation} 
Again, \eqref{ConvexIneq} holds for any $I$ with cdf $F_I\in \mathcal{I}\left(F_{I_1},F_{I_2}\right)$, i.e.\ for any dependence structure of $\left(I_1,I_2\right)$. 

The payoff function $r$ defined in \eqref{Eq1} is neither concave nor convex. It is a priori unclear whether extreme dependence structures lead to upper and lower bounds. In case they do, it is also unclear which of the two dependence structures (i.e. comonotonicity or countermonotonicity) would be the upper bound and which one would be the lower bound. The aim of this paper is to study when the order \eqref{Eq3-15} is either preserved or reversed after applying the transformation $r$ defined in \eqref{Eq1}.

\section{Dependence bounds} \label{Sec-Bounds}
Consider the function $r$ defined in \eqref{Eq1}, which can also be expressed as follows:
$$r:x\mapsto \frac{B}{\epsilon-\delta}\left\{\begin{array}{ll}0,& \qquad \text{for } x \leq \delta,\\ x-\delta, & \qquad \text{for } \delta \leq x \leq \epsilon,\\ \epsilon - \delta,& \qquad \text{for } \epsilon \leq x. \end{array}\right.$$ 
Throughout the paper, the points $\delta$ and $\epsilon$ satisfy $\delta<\epsilon$, and are omitted from the notation of the function $r$. The notation $B$ can be relevant in practice as it stands for the maximum payable amount, or the principal in case of a bond. However, for ease of notation, and without loss of generality, it is assumed that $B=\epsilon - \delta$.

This section aims at answering the central question of the paper. Namely, under what conditions do the expectations of $r(I^c)$ and $r(I^{cm})$ constitute bounds for the expectation of the original payoff $r(I)$? The answer provided in this paper is based on the result of the following lemma. A proof can be found in Appendix \ref{Theorem-proof}.
\begin{lemma}\label{Theorem}
	For any r.v.'s $X$ and $Y$ such that $X\preceq_{cx} Y$, the following statements hold:
	\begin{enumerate}
		\item If $\delta,\epsilon \in D^{(j)}_{\leq}(X,Y)$ for a given $j=0,1,\ldots,\frac{N-1}{2}$, then $\mathbb{E}\left[r(Y)\right] \leq \mathbb{E}\left[r(X)\right]$,
		\item If $\delta,\epsilon \in D^{(j)}_{\geq}(X,Y)$ for a given $j=0,1,\ldots,\frac{N-1}{2}$, then $\mathbb{E}\left[r(X)\right] \leq \mathbb{E}\left[r(Y)\right]$,
	\end{enumerate}
where $N$ is the number of crossing points of the cdf's of $X$ and $Y$, and the sets $D^{(j)}_{\leq}(X,Y)$ and $D^{(j)}_{\geq}(X,Y)$ are defined in \eqref{Set3} and \eqref{Set4}, respectively.
\end{lemma}

Let $N^c$ and $N^{cm}$ be the number of crossing points of the pair of cdf's $\left(F_I,F_{I^c}\right)$ and $\left(F_I,F_{I^{cm}}\right)$, respectively. Since $I^c\preceq_{cx}I\preceq_{cx}I^{cm}$, Lemma \ref{Theorem} leads to:
\begin{enumerate}
	\item If $\delta,\epsilon \in D^{(j)}_{\leq}(I,I^{cm})$ for a given $j=0,1,\ldots,\frac{N^{cm}-1}{2}$, then  $\mathbb{E}[r(I^{cm})] \leq \mathbb{E}[r(I)]$,
	\item If $\delta,\epsilon \in D^{(j)}_{\geq}(I,I^{cm})$ for a given $j=0,1,\ldots,\frac{N^{cm}-1}{2}$, then  $\mathbb{E}[r(I)] \leq \mathbb{E}[r(I^{cm})]$,
	\item If $\delta,\epsilon \in D^{(l)}_{\leq}(I^c,I)$ for a given $l=0,1,\ldots,\frac{N^{c}-1}{2}$, then $\mathbb{E}[r(I)] \leq \mathbb{E}[r(I^{c})]$,
	\item If $\delta,\epsilon \in D^{(l)}_{\geq}(I^c,I)$ for a given $l=0,1,\ldots,\frac{N^{c}-1}{2}$, then $\mathbb{E}[r(I^c)] \leq \mathbb{E}[r(I)]$.
\end{enumerate}

\begin{figure}[!h]\centering \includegraphics[scale=0.6]{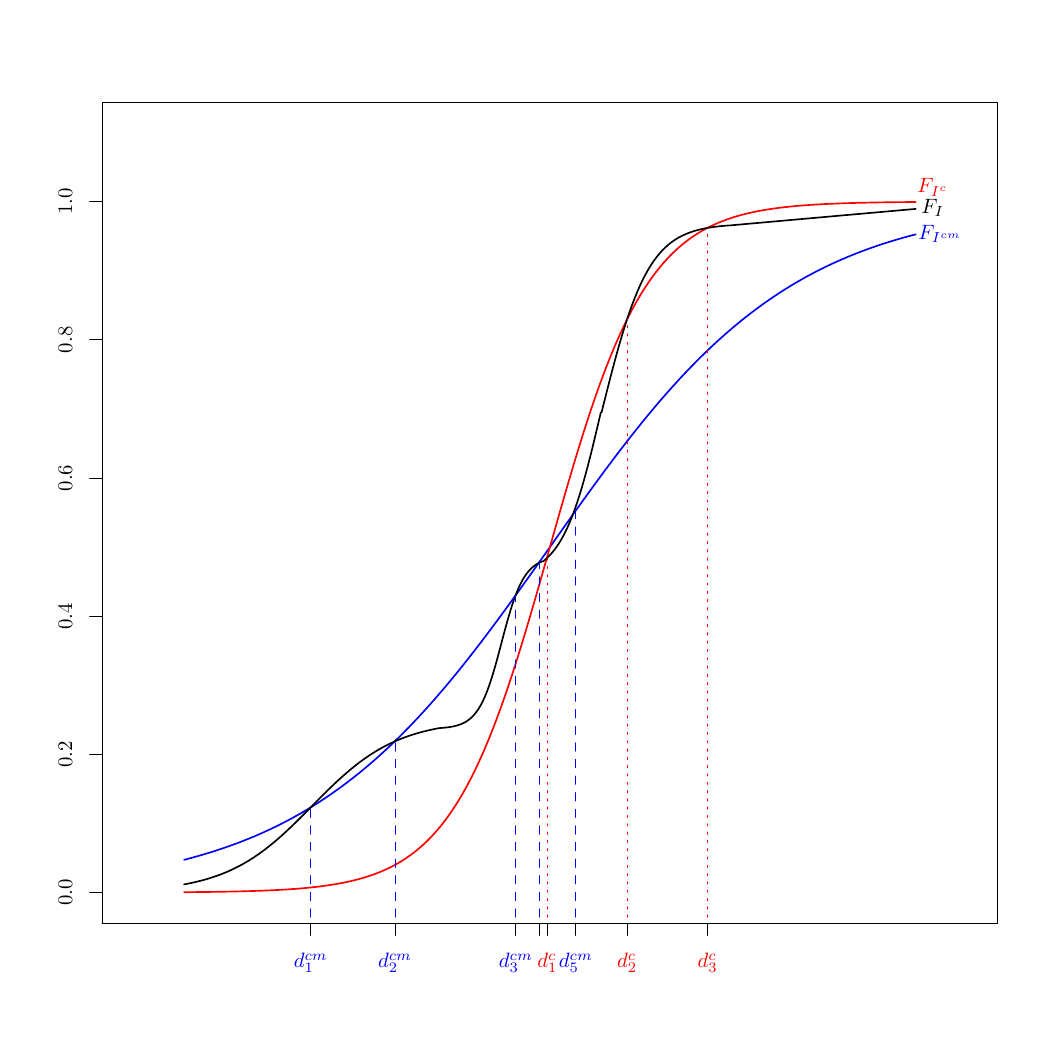}\caption{\small \it Illustration of the crossing points for arbitrarily chosen cdf's $F_I$, $F_{I^{c}}$ and $F_{I^{cm}}$.}\label{Figure0}
\end{figure}
Figure \ref{Figure0} illustrates the crossing points of the cdf's $F_I$, $F_{I^{c}}$ and $F_{I^{cm}}$, with $N^{cm}=5$ and $N^c=3$. The five crossing points of $F_I$ and $F_{I^{cm}}$ are denoted $d_j^{cm}$ for $j=1,\ldots,5$. The three crossing points of $F_I$ and $F_{I^c}$ are denoted $d_l^c$ for $l=1,2,3$. In this example, the inequality $\mathbb{E}\left[r(I^{cm})\right]\leq \mathbb{E}\left[r(I)\right]$ holds if $\delta<\epsilon\leq d_1^{cm}$, or $d_{2}^{cm}\leq \delta<\epsilon\leq d_{3}^{cm}$, or $d_{4}^{cm}\leq \delta<\epsilon \leq d_{5}^{cm}$. The inequality $\mathbb{E}\left[r(I)\right]\leq \mathbb{E}\left[r(I^{cm})\right]$ holds if $d_1^{cm}\leq\delta<\epsilon\leq d_{2}^{cm}$, or $d_{3}^{cm} \leq \delta<\epsilon\leq d_{4}^{cm}$, or $d_5^{cm}\leq \delta < \epsilon$. On the other hand, the inequality $\mathbb{E}\left[r(I)\right]\leq \mathbb{E}\left[r(I^c)\right]$ holds if $\delta < \epsilon \leq d_1^c$ or if $d_2^c\leq \delta < \epsilon \leq d^c_3$, whereas the inequality $\mathbb{E}\left[r(I^c)\right]\leq \mathbb{E}\left[r(I)\right]$ holds if $d_1^c\leq \delta <\epsilon \leq d_2^c$ or if $d_3^{c}\leq \delta <\epsilon$.

The remainder of this section is devoted to three important conclusions that can be drawn from the above inequalities and from Lemma \ref{Theorem}.
\subsection{The design of the payoff function $r$ determines the order}
The first and most obvious conclusion from Lemma \ref{Theorem} is that the order of the expected payoffs is not necessarily preserved nor reversed with the transformation $r$. As illustrated in Figure \ref{Figure0}, each of the extreme dependence structures can be either a lower bound or an upper bound. The order depends on the layers of the payoff function, i.e. on the values of $\delta$ and $\epsilon$. 

For the order of the expected payoffs to be clear, $\delta$ and $\epsilon$ must belong to the \textit{same} interval. For instance, in the example of Figure \ref{Figure0}, $\mathbb{E}\left[r(I)\right] \leq \mathbb{E}\left[r(I^{cm})\right]$ if $\delta,\epsilon \in \left[d_{1}^{cm},d_{2}^{cm}\right]$, or $\delta,\epsilon \in \left[d_{3}^{cm},d_{4}^{cm}\right]$, or $\delta,\epsilon \in [d_5^{cm},u_I)$. However, for $\delta\in \left[d_{1}^{cm},d_{2}^{cm}\right]$ and $\epsilon \in \left[d_{3}^{cm},d_{4}^{cm}\right]$, the inequality $\mathbb{E}\left[r(I)\right] \leq \mathbb{E}\left[r(I^{cm})\right]$ does not necessarily hold.

\subsection{Extreme dependence structures may bound from the same side}
A second conclusion from Lemma \ref{Theorem} is that it is even possible that both extreme dependence structures bound the expected payoff from the same side. If 
$$\delta,\epsilon \in \left(D^{(j)}_{\leq}\left(I,I^{cm}\right) \cap D^{(l)}_{\geq}\left(I^c,I\right)\right),$$
for some $j\in\{0,1,\ldots,\frac{N^{cm}-1}{2}\}$ and $l\in\{0,1,\ldots,\frac{N^c-1}{2}\}$ with 
$D^{(j)}_{\leq}\left(I,I^{cm}\right) \cap D^{(l)}_{\geq}\left(I^c,I\right) \neq \emptyset$, then: $$\mathbb{E}\left[r(I)\right]\geq \mathbb{E}\left[r(I^{cm})\right]\qquad \text{ and } \qquad \mathbb{E}\left[r(I)\right]\geq \mathbb{E}\left[r(I^c)\right].$$
Analogously, if 
$$\delta,\epsilon \in \left(D^{(j)}_{\geq}\left(I,I^{cm}\right) \cap D^{(l)}_{\leq}\left(I^c,I\right)\right),$$
for some $j\in\{0,1,\ldots,\frac{N^{cm}-1}{2}\}$ and $l\in\{0,1,\ldots,\frac{N^c-1}{2}\}$ with $D^{(j)}_{\geq}\left(I,I^{cm}\right) \cap D^{(l)}_{\leq}\left(I^c,I\right) \neq \emptyset$, then:
$$\mathbb{E}\left[r(I)\right]\leq \mathbb{E}\left[r(I^{cm})\right]\qquad \text{ and } \qquad \mathbb{E}\left[r(I)\right]\leq \mathbb{E}\left[r(I^c)\right].$$
Using the illustration from Figure \ref{Figure0}, if $d_1^c \leq \delta < \epsilon \leq d_5^{cm}$, then $\mathbb{E}\left[r(I)\right]\geq \mathbb{E}\left[r(I^{cm})\right]$ and $\mathbb{E}\left[r(I)\right]\geq \mathbb{E}\left[r(I^c)\right]$. For $d_1^{cm}\leq \delta < \epsilon \leq d_2^{cm}$, then $\mathbb{E}\left[r(I)\right]\leq \mathbb{E}\left[r(I^{cm})\right]$ and $\mathbb{E}\left[r(I)\right]\leq \mathbb{E}\left[r(I^c)\right]$.

In such cases, extreme dependence structures are less relevant in the context of dependence model risk. In particular, the dependence uncertainty spread is not obtained from the difference, nor the absolute difference, of the expected payoffs of the comonotonic and countermonotonic transforms.

It is worth noting that in the case of convex payoff functions, the order of the expectations reflects the preference and aversion towards an extreme dependence structure, which themselves are intuitively justified by diversification arguments. For instance, comonotonicity of $(I_1,I_2)$ (or equivalently, countermonotonicity of $(I_1,-I_2)$) leads to a lower bound for the expected stop-loss payoff of the difference $I_1-I_2$ because under this extreme dependence structure, the risk of $I_1$ is offset by that of $-I_2$. In the present situation with the payoff function $r$, the order of the expectations is highly dependent on the design of the payoff function.
\subsection{Dependence-dependent bounds}
A third, and more subtle, conclusion from Lemma \ref{Theorem} is that the conditions that unveil the order of the expected payoffs are specific to the dependence structure of the original indices $I_1$ and $I_2$. In other words, the sets $D_{\geq}(I,I^{cm})$, $D_{\leq}(I,I^{cm})$, $D_{\geq}(I^c,I)$ and $D_{\leq}(I^c,I)$, all are specific to the distribution of $I$, which itself is determined by the dependence structure of the random vector $\left(I_1,I_2\right)$. Thus, values of $\delta$ and $\epsilon$ that preserve/reverse the order for a given dependence structure of $\left(I_1,I_2\right)$ may not do so for another dependence structure. For instance, in Figure \ref{Figure0}, altering the dependence structure of $\left(I_1,I_2\right)$ would result in a different cdf $F_I$ that crosses $F_{I^c}$ and $F_{I^{cm}}$ at different points.

This observation highlights an important contrast with the case where the payoff is convex. In particular, for the expected stop-loss payoff and for any fixed $x\in\mathbb{R}$, the inequality \eqref{ConvexIneq} holds for \textit{any} random difference $I$ with cdf $F_I\in\mathcal{I}\left(F_{I_1},F_{I_2}\right)$, i.e.\ for any dependence structure of $\left(I_1,I_2\right)$. In contrast, Lemma \ref{Theorem} shows that for fixed $\delta$ and $\epsilon$, the inequalities after the transformation $r$ are specific to a random difference with fixed $F_I\in\mathcal{I}\left(F_{I_1},F_{I_2}\right)$, and do not necessarily hold for other modifications of $I$ from the Fr\'{e}chet space $\mathcal{F}\left(F_{I_1},F_{I_2}\right)$.

The consequence is that if the bounds themselves depend on the dependence structure, then in some cases, they do not play the role of dependence model risk safeguards anymore. Nevertheless, if the crossing points are unrelated to the dependence structure of $\left(I_1,I_2\right)$ for some given marginal distributions, then using $r(I^c)$ and $r(I^{cm})$ in the context of dependence model risk would be meaningful. This aspect is discussed in the next section.

\subsection{Summary of the main results on the bounds}
The goal of this section was to determine conditions under which the expectations of $r(I^c)$ and $r(I^{cm})$ constitute bounds for the expectation of the original payoff $r(I)$. The conclusion is that $r(I^c)$ and $r(I^{cm})$ lead to bounds if:
$$\delta, \epsilon \in \left(D^{(j)}_{\geq}\left(I,I^{cm}\right)\cap D^{(l)}_{\geq}\left(I^c,I\right)\right),$$
for some $j\in\{0,1,\ldots,\frac{N^{cm}-1}{2}\}$ and $l\in\{0,1,\ldots,\frac{N^c-1}{2}\}$ such that $D^{(j)}_{\geq}\left(I,I^{cm}\right)\cap D^{(l)}_{\geq}\left(I^c,I\right)\neq \emptyset$, or if:
$$\delta, \epsilon \in \left(D^{(j)}_{\leq}\left(I,I^{cm}\right)\cap D^{(l)}_{\leq}\left(I^c,I\right)\right),$$
for some $j\in\{0,1,\ldots,\frac{N^{cm}-1}{2}\}$ and $l\in\{0,1,\ldots,\frac{N^c-1}{2}\}$ such that $D^{(j)}_{\leq}\left(I,I^{cm}\right)\cap D^{(l)}_{\leq}\left(I^c,I\right)\neq \emptyset$.
In particular, $\mathbb{E}\left[r(I^c)\right] \leq \mathbb{E}\left[r(I)\right] \leq \mathbb{E}\left[r(I^{cm})\right]$ in the former case, whereas $\mathbb{E}\left[r(I^{cm})\right] \leq \mathbb{E}\left[r(I)\right] \leq \mathbb{E}\left[r(I^{c})\right]$ in the latter. Thus, the transformation $r$ preserves the inequality $I^c \preceq_{cx} I \preceq_{cx} I^{cm}$ under the former condition, and reserves it under the latter condition. However, for fixed $\delta$ and $\epsilon$, these relationships do not necessarily hold for any dependence structure of $\left(I_1,I_2\right)$, and depend on the location of the crossing points of the cdf's $F_I$, $F_{I^{c}}$ and $F_{I^{cm}}$.

\section{Crossing points}\label{Sec-Crossing}
The results of Section \ref{Sec-Bounds} show that it is important to study the crossing points of the cdf's of $I$, $I^c$ and $I^{cm}$ in order to determine the order of their expected stop-loss payoffs under the transformation $r$. This section discusses some properties of the crossing points. In the remainder of the paper, the `\textit{pairwise crossing points}' refer to the crossing points of the three pairs of cdf's $(F_I,F_{I^c})$, $(F_I,F_{I^{cm}})$ and $(F_{I^c},F_{I^{cm}})$, and the `\textit{uniqueness of the pairwise crossing points}' refers to the situation where each of these pairs of cdf's has a unique crossing point.

\subsection{Uniqueness of the pairwise crossing points}
A convenient situation is when $F_I$, $F_{I^c}$ and $F_{I^{cm}}$ have unique pairwise crossing points, and that these crossing points all coincide, i.e.\ for any $F_I\in \mathcal{I}\left(F_{I_1},F_{I_2}\right)$ such that $I$ does not correspond to a comonotonic or countermonotonic transform, there exists a unique $d$ satisfying $F_I(d)=F_{I^c}(d)=F_{I^{cm}}(d)$, with:
\begin{eqnarray*}F_{I^{c}}(x)\leq F_{I}(x)\leq F_{I^{cm}}(x), &&\quad \text{for \ \ \ } x\leq d,\\
	F_{I^{cm}}(x)\leq F_{I}(x)\leq F_{I^{c}}(x), &&\quad \text{for \ \ \ }  d\leq x.\end{eqnarray*}
In this case, the transformation $r$ preserves the inequality \eqref{ConvexIneq} if $d\leq \delta< \epsilon$, and reverses it if $\delta< \epsilon\leq d$. Extreme dependence structures are therefore appropriate bounds, provided $\delta$ and $\epsilon$ are both larger than $d$, or both smaller than $d$.

Another interesting situation with a milder condition is when the pairwise crossing points of $F_I$, $F_{I^c}$ and $F_{I^{cm}}$ are unique, but do not necessarily coincide. Let $d^c$, $d^{cm}$ and $d^{\star}$ be those unique points, with  $D(I^c,I)=\{d^c\}$, $ D(I,I^{cm})=\{d^{cm}\}$ and $ D(I^c,I^{cm})=\{d^{\star}\}$, such that:
$$\begin{array}{lll}
	F_{I}\left(d^c\right) = F_{I^c}\left(d^c\right), & \quad F_{I}\left(d^{cm}\right) = F_{I^{cm}}\left(d^{cm}\right), & \quad 	F_{I^c}\left(d^{\star}\right) = F_{I^{cm}}\left(d^{\star}\right).\end{array}$$
In particular, $N^c=N^{cm}=N^{\star}=1$, where $N^c$, $N^{cm}$ and $N^{\star}$ are the numbers of elements in $D(I^c,I)$, $D(I,I^{cm})$ and $D(I^c,I^{cm})$, respectively. From Lemma \ref{Theorem}, it follows that:
\begin{enumerate}
	\item If $\max\{d^{cm}, d^c\}\leq \delta <\epsilon$, then $\mathbb{E}\left[r(I^c)\right] \leq \mathbb{E}\left[r(I)\right] \leq \mathbb{E}\left[r(I^{cm})\right]$,
	\item If $\delta < \epsilon \leq \min\{d^{cm}, d^c\}$, then $\mathbb{E}\left[r(I^{cm})\right] \leq \mathbb{E}\left[r(I)\right] \leq \mathbb{E}\left[r(I^{c})\right]$.
\end{enumerate}
In this case, the transformation $r$ leads to an order which is consistent with $I^c \preceq_{cx} I \preceq_{cx} I^{cm}$ if $\delta$ and $\epsilon$ are both larger than the largest of $d^c$ and $d^{cm}$, and reverses that order if both are smaller than the smallest of $d^c$ and $d^{cm}$. 

Analytical results on the crossing points of cdf's are generally hard to derive, and even determining the number of crossing points is not straightforward. In particular, it is hard to verify analytically whether these situations where $N^c=N^{cm}=N^{\star}=1$ are common. The following link is provided as an alternative to explore graphically the number of crossing points:
\begin{center}\hyperref{https://hamzahanbali.shinyapps.io/crossingpoints/}{}{}{\textcolor{blue}{\underline{Crossing points for some continuous distribution functions}}}\end{center}
The link contains a wide palette of marginal distributions and copula models. The graphical observation shows that in many cases, the three cdf's have a unique crossing point, i.e. that $N^c=N^{cm}=N^{\star}=1$. This suggests the tempting conjecture that for continuous and unimodal marginal distributions, there exist unique pairwise crossing points between $F_I$, $F_{I^c}$ and $F_{I^{cm}}$ when $I$ is neither a comonotonic nor a countermonotonic transform. But despite the fact that this property holds for many combinations available in the above link, the conjecture fails in some cases. For instance, the cdf's $F_I$ and $F_{I^{cm}}$ have three crossing points when $I_1$ and $I_2$ are independent Variance Gamma distributed with respective location parameters $0$ and $-4$, spread parameters $2$ and $12$, asymmetry parameters $-4$ and $-3$, and shape parameters $3$ and $12$. It is also worth noting that in general, the sets $D_{\leq}^{(j)}$ and $D_{\geq}^{(i)}$ can be even more complex for some marginal distributions. For instance, \cite{KaasGoovaertsTang2004} constructed a univariate distribution such that the cdf of $I$ crosses infinitely many times with that of $I^{cm}$. Multiple crossing points can also occur when the marginal distributions are multi-modal.

The remainder of this section focuses on the case where the numbers of pairwise crossing points are unique, i.e. $N^c=N^{cm}=N^{\star}=1$. As the graphical observation from the above link shows, this case covers many combinations of marginal distributions and dependence structures, and is therefore relevant.

\subsection{Order of the crossing points}
The following theorem provides an order of the crossing points in case of unique pairwise crossing points for general r.v.'s; see Appendix \ref{Theorem_crossing-proof} for a proof.
\begin{theorem}\label{Theorem_crossing}
	Consider the r.v.'s $X$, $Y$ and $Z$ with distinct cdf's such that $X\preceq_{cx}Y\preceq_{cx}Z$. Let $d_{XY}$, $d_{XZ}$ and $d_{YZ}$ be the unique pairwise crossing points such that $F_{X}\left(d_{XY}\right)=F_{Y}\left(d_{XY}\right)$, $F_{X}\left(d_{XZ}\right)=F_{Z}\left(d_{XZ}\right)$, and  $F_{Y}\left(d_{YZ}\right)=F_{Z}\left(d_{YZ}\right)$. Then one of the following conditions hold:
	\begin{enumerate}
		\item $d_{XY}<d_{XZ}<d_{YZ}$, or
		\item $d_{YZ}<d_{XZ}<d_{XY}$, or
		\item $d_{XY}=d_{XZ}=d_{YZ}$.
	\end{enumerate}
\end{theorem}

Theorem \ref{Theorem_crossing} shows that if two crossing points are equal, then all three crossing points are equal. Otherwise, if two crossing points are different, then the crossing point of the cdf's of the countermonotonic and comonotonic differences must be in the middle. Specifically, since $I^c \preceq_{cx} I \preceq_{cx} I^{cm}$, then one of the following conditions hold:
	\begin{enumerate}
	\item $d^c<d^{\star}<d^{cm}$, or
	\item $d^{cm}<d^{\star}<d^c$, or
	\item $d^c=d^{\star}=d^{cm}$.
\end{enumerate}

The following corollary is an immediate consequence of Theorem \ref{Theorem_crossing}. The proof is omitted.
\begin{corollary}\label{Corollary1}
	The cdf $F_{I^{cm}}$ has a unique crossing point $d$ with any other cdf from $\mathcal{I}\left(F_{I_1},F_{I_2}\right)$ if and only if any two distinct cdf's from $\mathcal{I}\left(F_{I_1},F_{I_2}\right)$ have a unique crossing point $d$.
\end{corollary}

Corollary \ref{Corollary1} means that in order to verify whether the pairwise crossing points of all distinct cdf's in $\mathcal{I}\left(F_{I_1},F_{I_2}\right)$ coincide, it is equivalent to prove that for any cdf $F_I\in\mathcal{I}\left(F_{I_1},F_{I_2}\right)$ which is not constructed from a countermonotonic modification, the solution to the equation $F_{I^{cm}}(z) = F_{I}(z)$ is unique and unrelated to the dependence structure of $\left(I_1,I_2\right)$. Alternatively, it is equivalent to prove that the function
$$ z \mapsto F^{-1}_{I_1}\left(F_I(z)\right) - F^{-1}_{I_2}\left(1-F_{I}(z)\right)$$
has a unique fixed point which is unrelated to the dependence structure of $\left(I_1,I_2\right)$. Note that the corollary is valid for other cdf's from $\mathcal{I}\left(F_{I_1},F_{I_2}\right)$, but taking $F_{I^{cm}}$ as the reference cdf is more convenient because the inverse cdf $F^{-1}_{I^{cm}}$ can be expressed as the difference of the marginal quantiles \citep{Dhaene2002a,HanbaliLinders}.

\subsection{Crossing points and dispersive order}
In general, the r.v.\ $Y$ is said to be greater than $X$ in the dispersive order, with the notation $X\preceq_{DISP} Y$, if and only if $p\mapsto F^{-1}_{Y}(p)-F^{-1}_{X}(p)$ is increasing in $p$. Dispersive order of $I_1$ and $I_2$ leads to an interesting result regarding the crossing points of $I^c$ and $I^{cm}$. The proof of the following theorem can be found in Appendix \ref{Theorem_Disp-proof}.
\begin{theorem}
	\label{Theorem_Disp}
	If $I_1 \preceq_{DISP} I_2$ or $I_2 \preceq_{DISP} I_1$, then there exists a unique crossing point $d^{\star}$ which follows from:
	$$F_{I^c}(d^{\star}) = F_{I^{cm}}(d^{\star}) = \frac{1}{2}.$$
\end{theorem}
This theorem provides a condition under which the crossing point of $F_{I^c}$ and $F_{I^{cm}}$ is unique and easy to determine from the marginal distributions. In particular, the crossing point $d^{\star}$ corresponds to the difference of the marginal medians, i.e.\
\begin{equation}\label{SimpleCrossing}d^{\star}=F^{-1}_{I^{cm}}\left(\frac{1}{2}\right) = F^{-1}_{I_1}\left(\frac{1}{2}\right) - F^{-1}_{I_2}\left(\frac{1}{2}\right).\end{equation}

Note however that this result does not guarantee that the pairwise crossing points with the other cdf's from $\mathcal{I}\left(F_{I_1},F_{I_2}\right)$ are also unique, nor that the median is one of them.

\subsection{Crossing points under conditions of symmetry}
The proof of the following theorem can be found in Appendix \ref{Theorem_Sym-proof}.
\begin{theorem}\label{Theorem_Sym}
	Let $I_i \overset{d}{=} \mu_i + \sigma_i W_i$, where $W_1$ and $W_2$ are symmetric about $0$ and identically distributed with cdf $F_W$. Consider the radially symmetric copula $C$, i.e.\ its density satisfies $c(u,v)=c(1-u,1-v)$ for all $\left(u,v\right)\in\left[0,1\right]^2$. Then for any random difference $I$ with cdf $F_I\in \mathcal{I}\left(F_{I_1},F_{I_2}\right)$ such that $\left(W_1,W_2\right)$ has a copula $C$ which is neither comonotonic nor countermonotonic, all three cdf's $F_I$, $F_{I^c}$ and $F_{I^{cm}}$ have a unique crossing point, such that:
	$$d^{c}=d^{cm}=d^{\star}=\mu_1-\mu_2.$$
\end{theorem}

The unique crossing point of the three cdf's corresponds again to the median of the distributions. Since $d^{c}=d^{cm}=d^{\star}$, extreme dependence structures are always appropriate bounds if $d\leq \delta<\epsilon$ or $\delta<\epsilon\leq d$. Therefore, under the conditions of Theorem \ref{Theorem_Sym}, the question of whether the transformation $r$ preserves or reverses the inequalities is simple to answer.

\section{Application to longevity trend bonds}\label{Sec-NumericalAnalysis}
In the previous section, it was shown that the uniqueness of the pairwise crossing points leads to a convenient situation where the order of the expectations can easily be determined. Further, Theorems \ref{Theorem_Disp} and \ref{Theorem_Sym} provide conditions under which the unique crossing point simply corresponds to the difference of the marginal medians. 

The present section investigates the uniqueness of the crossing point and its consequence on the order of expected payoffs in a practical context, which is that of mortality-linked securities. The analysis focuses on longevity trend bonds, whose payoff function is given by $r(I)$, where $r$ is as defined in \eqref{Eq1} and the underlying $I$ is a difference of two mortality indices $I_1$ and $I_2$. 

Using the construction from Swiss Re's 2010 Kortis bond for the indices $I_1$ and $I_2$, the crossing points of the cdf's $F_I$, $F_{I^c}$ and $F_{I^{cm}}$ are investigated in order to determine the sets $D^{(j)}_{\leq}(I,I^{cm})$, $D^{(j)}_{\geq}(I,I^{cm})$, $D^{(l)}_{\leq}(I^c,I)$ and $D^{(l)}_{\geq}(I^c,I)$, and hence provide insights into the values of $\delta$ and $\epsilon$ leading to $\mathbb{E}\left[r(I^c)\right] \leq \mathbb{E}\left[r(I)\right] \leq \mathbb{E}\left[r(I^{cm})\right]$ or $\mathbb{E}\left[r(I^{cm})\right] \leq \mathbb{E}\left[r(I)\right] \leq \mathbb{E}\left[r(I^{c})\right]$. In particular, a large simulation study using real mortality data is conducted to verify whether the uniqueness of the crossing point holds. Where the uniqueness of the crossing point holds, the study investigates whether approximating that point with the median is appropriate.

Subsection \ref{Sec51} contains some background on longevity trend bonds. Subsection \ref{Sec52} presents the methodology used in the numerical study. Subsection \ref{Sec53} reports the results on the crossing points, whereas Subsection \ref{Sec54} reports the results on the expected payoffs. Subsection \ref{Sec55} provides a brief summary of the numerical study.
\subsection{Background}\label{Sec51}
Mortality-linked securities refer to risk-transfer instruments from the emerging life market \citep{Blake2006,Blake2013,BiffisBlake2014,BlakeIME2017,Blake2018}. They provide alternative hedging solutions to institutions exposed to systematic longevity or mortality risks. The development of this market requires the pricing and the risk assessment of the hedging instruments it offers. Issuers face two salient challenges, namely, increasing the likelihood of a successful launch and achieving a good hedge. Carrying out a sound analysis of the payoff of longevity-linked instruments is a crucial task in order to address these challenges. For many mortality-linked securities, the analysis of the payoff requires modeling the dependence between the mortality improvements in multiple populations. However, as pointed out in the comparative studies of \cite{Enchev2017} and \cite{Villegas2017}, multi-population models can produce different forecasts despite their close in-sample performance, thereby raising the issue of dependence model risk. Thus, in this context, it is important to study whether extreme dependence structures of the underlying mortality indices can be used to address dependence model risk.

The longevity trend bonds of interest gained popularity in the literature after the launch in 2010 of Swiss Re's Kortis bond \citep{HuntBlake2015,ChenMacMinnSun2015,ChenMacMinnSun2017,LiTang2019}. Their payoff is a function of the spread between the mortality improvements in two different populations at maturity. The spread at maturity, denoted by $I$ and referred to as the \textit{longevity divergence index}, captures the divergence of mortality improvements of two populations, such that $I=I_1-I_2$ where $I_i$ is the index measuring the mortality improvement at maturity in population $i$ \citep{HuntBlake2015,LiTang2019}. The payoff to the issuer of the longevity trend bond is given by $r(I)$, where $r$ is defined in \eqref{Eq1}. The point $\delta$ is referred to as the attachment point, below which the bondholders receive the
full principal $B$ back, whereas the point $\epsilon$ is referred to as the exhaustion point, beyond which the principal goes to the issuer. Again, the simplifying assumption $B=\epsilon-\delta$ is adopted in this analysis.

Let $\mu_{x,t}^{(i)}$ be the crude mortality rate at age $x$ in year $t$ for population $i\in\{1,2\}$. Using the construction from Swiss Re's Kortis bond, the index $I_i$ at the maturity date $t_0+T$ is defined as follows:
\begin{equation}\label{Indexi}
	I_i = 1 - \frac{1}{\omega^{(i)}-\alpha^{(i)} + 1}\underset{x=\alpha^{(i)}}{\overset{\omega^{(i)}}{\sum}} \left(\frac{\mu_{x,t_0+T}^{(i)}}{\mu_{x,t_0}^{(i)}}\right)^{\frac{1}{T}},
\end{equation}
where $[\alpha^{(i)},\omega^{(i)}]$ is the age range used for the index $I_i$, with $\alpha^{(i)}<\omega^{(i)}$.

The Swiss Re Kortis bond was intended to hedge against Swiss Re's exposure to basis risk between the UK annuity business and the US life assurance business. In particular, $I_1$ is the index for the English and Welsh population, whereas $I_2$ is the index for the US population. The averaging in \eqref{Indexi} was over ages 75-85 for the English and Welsh index, and 55-65 for the US index. The choice of these ages is justified by the fact that annuity holders are typically older than life assurance holders. The indices in the Kortis bond measure the improvement between 2008 and 2016, which gives a total of $T=8$ years. The attachment and exhaustion points are $\delta = 3.4\%$ and $\epsilon=3.9\%$. Further details on the Swiss Re Kortis bond can be found in \cite{HuntBlake2015}.

\subsection{Methodology}\label{Sec52}
The study uses six models for the marginal distributions $F_{I_1}$ and $F_{I_2}$. The first two models are the single-population models proposed in \cite{CBD2006} (henceforth CBD) and \cite{LeeCarter1992}. The third and the fourth models are the two-population models proposed in \cite{LiLee2005} as an extension of the Lee-Carter model, and the Common age effect (CAE) model proposed in \cite{Kleinow2015}. The fifth and the sixth models are based on the historical time series of the averages in \eqref{Indexi}, such that $I$ is either a difference of normal or log-normal r.v.'s; see Appendix \ref{AppendixModels} for more details on the marginal models.

Each of the six models specifies different marginal distributions. For a given age range $[\alpha^{(i)},\omega^{(i)}]$ and a given maturity $T$, the indices $I_1$ and $I_2$ are simulated under each of these assumptions. The $100,000$ simulations are then re-ordered using the rank of the observations to produce the longevity divergence index $I$ under different dependence structures, including comonotonicity and countermonotonicity. The re-ordering of the simulations based on the rank allows to change the dependence structure while keeping the same set of simulated values. The cdf of $I$ is then estimated numerically for each dependence structure. Note that since the Li-Lee and the CAE models are two-population models, they already specify a dependence structure. However, in the simulations, only the marginal distributions from these models are used in order to have more flexibility in the specification of the dependence structure.

The study is performed using data from 21 countries, and accounts for different combinations of countries, marginal models, copula models, age groups and maturities. The results are qualitatively comparable across virtually all combinations. In the next subsection, only the results of the 2010 Kortis bond are reported, with English and Welsh mortality data for $I_1$ and US mortality data for $I_2$, and with the age groups and maturity described above. The results for all other combinations are reported in the following link:
\begin{center}\hyperref{https://hamzahanbali.shinyapps.io/longevitydivergenceindex_crossingpoints/}{}{}{\textcolor{blue}{\underline{Distribution of the longevity divergence index and crossing points}}}\end{center}
This online tool displays the cdf's $F_{I}$, $F_{I^c}$ and $F_{I^{cm}}$, where $I$ is simulated from one of the six marginal models described above\footnote{Note that refreshing the web page causes new simulations to be generated, and leads to new values, but the differences are small.}. The tool also provides estimates of the crossing points, as well as the quantiles at the levels $0.05$, $0.5$ and $0.95$. The median $F^{-1}(0.5)$ allows to evaluate whether it is a good approximation for the crossing points. The lower and upper quantiles $F^{-1}(0.05)$ and $F^{-1}(0.95)$ inform on the quality of the approximation, and in particular, on whether the discrepancies between the crossing points and the median are significant compared to the support of the longevity divergence index.

All these values can be determined for different combinations of the indices $I_1$ and $I_2$. The first input is the country $i$. The second input is the age group determined by $\alpha^{(i)}$ and $\omega^{(i)}$. The third input is the marginal model. Unlike in the results reported in the core of the paper where the distributions of $I_1$ and $I_2$ are derived from the same model, the online tool allows to select different marginal models. Flexibility on the choice of the copula model is included too. The online tool also allows to select different maturities, attachment and exhaustion points.

\subsection{Results on the crossing points using English-Welsh and US data}\label{Sec53}
\begin{figure}[!h]
	\centering \includegraphics[width=1.02\textwidth]{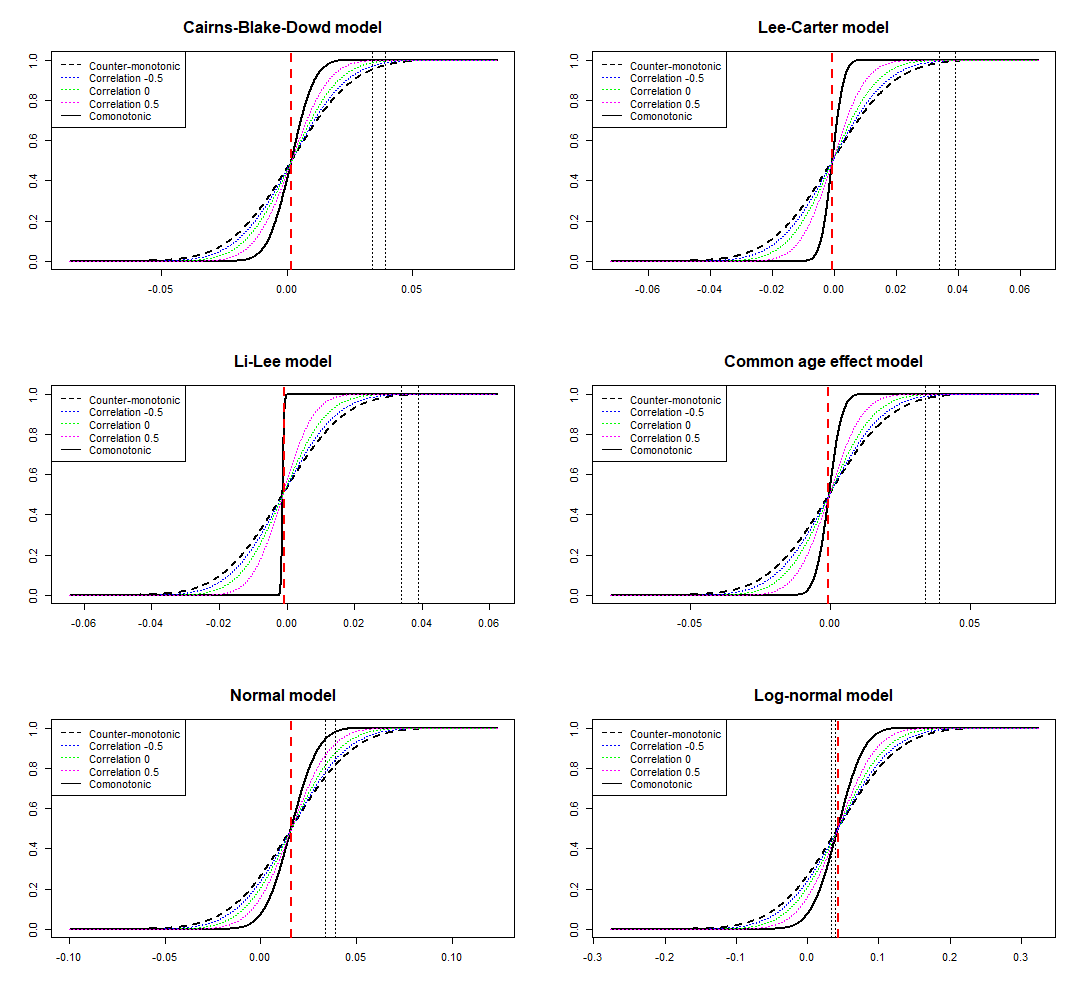}\caption{\it \small \textbf{Cdf's with  Gaussian copula} -- Cumulative distribution functions of the longevity divergence index $I$ under six different marginal distributions. Each panel corresponds to one of the six marginal distribution models, where both $I_1$ and $I_2$ follow the same model with different parameters. Within each panel, the thick black curve corresponds to comonotonicity (i.e.\ $F_{I^c}$) and the thick dashed black curve corresponds to countermonotonicity (i.e.\ $F_{I^{cm}}$). The remaining dotted curves correspond to situations where $(I_1,I_2)$ has a correlation of $-0.5$, $0$ or $0.5$. The vertical dashed red line is the median of $I^{cm}$. The vertical dotted lines are the attachment and exhaustion points $\delta = 3.4\%$ and $\epsilon=3.9\%$.}\label{Figure1}
\end{figure}

\begin{figure}[!h]
	\centering \includegraphics[width=1.02\textwidth]{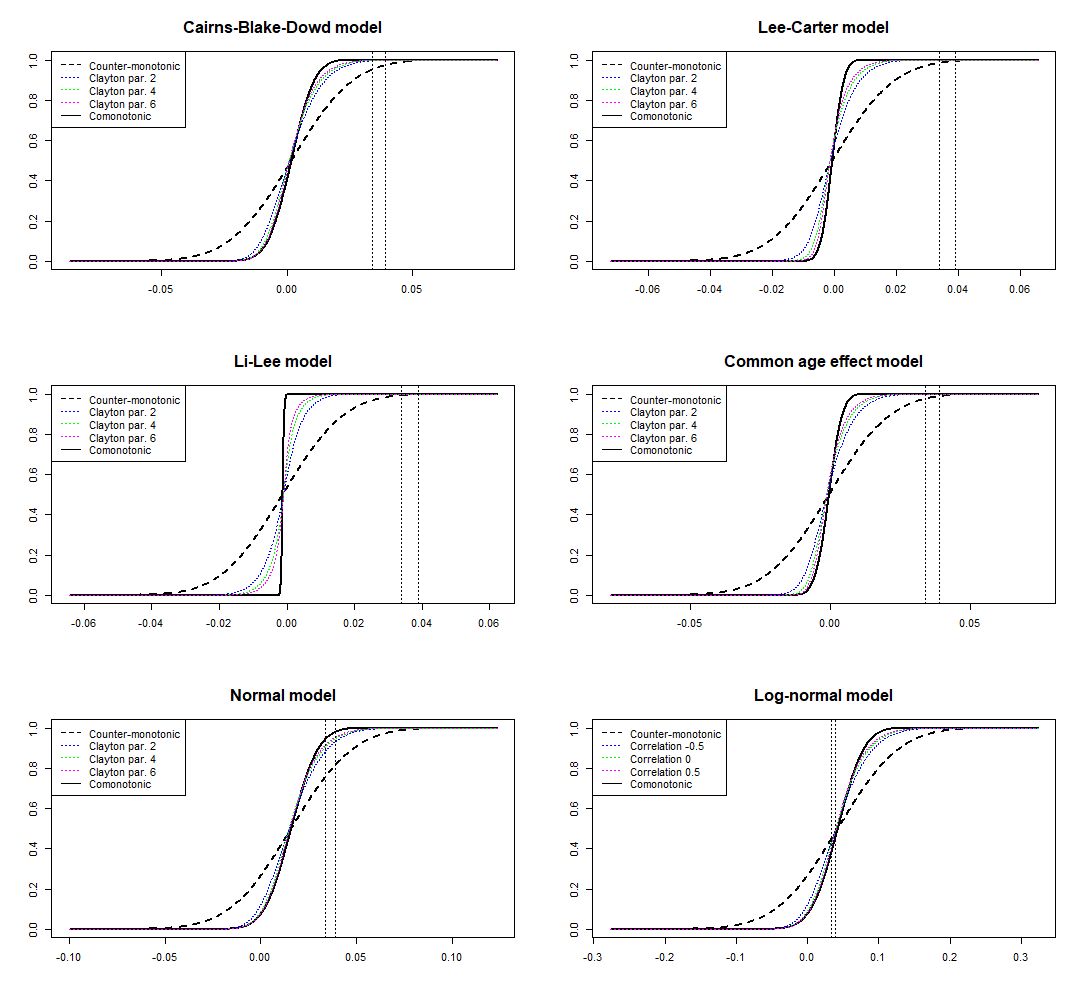}\caption{\it \small \textbf{Cdf's with Clayton copula} -- Cumulative distribution functions of the longevity divergence index $I$ under six different marginal distributions. Each panel corresponds to one of the six marginal distribution models, where both $I_1$ and $I_2$ follow the same model with different parameters. Within each panel, the thick black curve corresponds to comonotonicity (i.e.\ $F_{I^c}$) and the thick dashed black curve corresponds to countermonotonicity (i.e.\ $F_{I^{cm}}$). The remaining dotted curves correspond to situations where $(I_1,I_2)$ has a Clayton copula with parameter $2$, $4$ or $6$. The vertical dotted lines are the attachment and exhaustion points $\delta = 3.4\%$ and $\epsilon=3.9\%$.}\label{Figure2}
\end{figure}

Figure \ref{Figure1} displays the cdf's of the longevity divergence index $I$ under all marginal distribution models with a  Gaussian copula. Each panel corresponds to one of the six marginal distribution models, where both $I_1$ and $I_2$ follow the same model with different parameters. Within each panel, the thick black curve corresponds to comonotonicity (i.e.\ $F_{I^c}$) and the thick dashed black curve corresponds to countermonotonicity (i.e.\ $F_{I^{cm}}$). The remaining dotted curves correspond to situations where $(I_1,I_2)$ has a correlation of $-0.5$, $0$ or $0.5$. The vertical dashed red line is the median of $I^{cm}$. The vertical dotted lines are the attachment and exhaustion points $\delta = 3.4\%$ and $\epsilon=3.9\%$. Figure \ref{Figure2} displays the same cdf's but with a Clayton copula. The thick and dashed black curves still correspond to comonotonicity and countermonotonicity, but the remaining dotted curves correspond to situations where $(I_1,I_2)$ has a Clayton copula with parameter $2$, $4$ or $6$. The median of $I^{cm}$ is not displayed in Figure \ref{Figure2}. Table \ref{Table1} reports the numerical values of the medians of $I^{cm}$, $I^c$ and $I$, as well as the crossing points $d^{cm}$, $d^c$ and $d^{\star}$ for all marginal and dependence models.

The first observation from both Figures \ref{Figure1} and \ref{Figure2} is that all cdf's have single pairwise crossing points. Under the  Gaussian copula (Figure \ref{Figure1}), all crossing points appear to coincide with the median of $I^{cm}$. For the normal model (bottom-left), the result is not surprising because the marginal distributions and the copula satisfy the assumption of symmetry of Theorem \ref{Theorem_Sym}. On the other hand, under the Clayton copula (Figure \ref{Figure2}), the pairwise crossing points do not coincide, except for the Li-Lee model where the points appear to be close. The values reported in Table \ref{Table1} confirm these observations. For each model of the marginal distributions, the differences between all the medians and the crossing points are small under the Gaussian copula\footnote{The differences in Table \ref{Table1} under the Gaussian copula are mostly due to numerical errors. Indeed, the equality between the medians and the crossing points for normal marginal distributions with a Gaussian copula is theoretically exact, whereas small differences are observed in the table. It is worth mentioning that the differences in the model with normal marginal distributions and a Gaussian copula are comparable to those in the other models with the same copula.}, but they are comparatively larger under the Clayton copula.   

The second observation from both figures is that for all marginal models except the log-normal model, the payoff function $r$ leads to an order of expected payoffs which is consistent with the inequality $I^c \preceq_{cx} I \preceq_{cx} I^{cm}$ for the choice of attachment and exhaustion points $\delta = 3.4\%$ and $\epsilon=3.9\%$. In particular, for the CBD, the Lee-Carter, the Li-Lee, the CAE, and the normal models, $\delta = 3.4\%$ and $\epsilon=3.9\%$ are both larger than the crossing points under all dependence specifications, and hence, $\mathbb{E}\left[r(I^c)\right] \leq \mathbb{E}\left[r(I)\right] \leq \mathbb{E}\left[r(I^{cm})\right]$. 

For the log-normal model under the Gaussian copula, the (unique and approximately coinciding) crossing point is above both $\delta$ and $\epsilon$, meaning that the order is reversed, i.e.\ $\mathbb{E}\left[r(I^{cm})\right] \leq \mathbb{E}\left[r(I)\right] \leq \mathbb{E}\left[r(I^{c})\right]$. But under the Clayton copula, the log-normal model leads to the ambiguous situation where:
$$\delta<d^{cm} < \epsilon < d^c.$$
In particular, based on the results of the present paper for the log-normal model under the Clayton copula, it is only possible to ascertain that the comonotonic transform leads to an upper bound.

\begin{table}[!h]
	\centering\footnotesize
	\begin{tabular}{l|ccc|ccc}
		\toprule
		& med$(I^{cm})$&med$(I^{c})$&med$(I)$&$d^{\star}$ &$d^c$ &$d^{cm}$	\\
		\cline{1-7}
		
		\textit{Cairns-Blake-Dowd model}  &&&&&&\\
		&  0.001729&  0.001729&          &  0.001729&          &          \\
		\ \ \ \ \ \ \ Cor$(I_1,I_2)=-0.5$&          &          &  0.001744&          &  0.001728&  0.001843\\
		\ \ \ \ \ \ \ Cor$(I_1,I_2)=0$&          &          &  0.001766&          &  0.001708&  0.001876\\
		\ \ \ \ \ \ \ Cor$(I_1,I_2)=0.5$&          &          &  0.001783&          &  0.001680&  0.001865\\
		\ \ \ \ \ \ \ Clayton param. $2$        &          &   & 0.000701       &    & 0.004174 &-0.000367\\
		\ \ \ \ \ \ \ Clayton param. $4$        &          &   & 0.000709       &    & 0.005987 &-0.000080\\
		\ \ \ \ \ \ \ Clayton param. $6$        &          &   & 0.000888       &    & 0.001122 & 0.000365\\
		\cline{1-7}
		\textit{Lee-Carter model} &&&&&&\\
		& -0.000632& -0.000631&          & -0.000631&          &          \\
		\ \ \ \ \ \ \ Cor$(I_1,I_2)=-0.5$&          &          & -0.000617&          & -0.000635& -0.000614\\
		\ \ \ \ \ \ \ Cor$(I_1,I_2)=0$&          &          & -0.000589&          & -0.000645& -0.000503\\
		\ \ \ \ \ \ \ Cor$(I_1,I_2)=0.5$&          &          & -0.000601&          & -0.000645& -0.000567\\
		\ \ \ \ \ \ \ Clayton param. $2$        &          &   &-0.001357       &    & 0.000020 &-0.001815\\
		\ \ \ \ \ \ \ Clayton param. $4$        &          &   &-0.001459       &    & 0.000559 &-0.001793\\
		\ \ \ \ \ \ \ Clayton param. $6$        &          &   &-0.001407       &    & 0.000921 &-0.001646\\
		\cline{1-7}
		\textit{Li-Lee model} &&&&&&\\
		& -0.001196& -0.001196&          & -0.001196&          &          \\
		\ \ \ \ \ \ \ Cor$(I_1,I_2)=-0.5$&          &          & -0.001181&          & -0.001196& -0.001055\\
		\ \ \ \ \ \ \ Cor$(I_1,I_2)=0$&          &          & -0.001196&          & -0.001196& -0.001196\\
		\ \ \ \ \ \ \ Cor$(I_1,I_2)=0.5$&          &          & -0.001175&          & -0.001197& -0.001151\\
		\ \ \ \ \ \ \ Clayton param. $2$        &          &   &-0.001044       &    &-0.001251 &-0.000957\\
		\ \ \ \ \ \ \ Clayton param. $4$        &          &   &-0.001007       &    &-0.001256 &-0.000951\\
		\ \ \ \ \ \ \ Clayton param. $6$        &          &   &-0.000979       &    &-0.001273 &-0.000938\\
		\cline{1-7}
		\textit{Common age effect model} &&&&&&\\
		& -0.000254& -0.000255&          & -0.000255&          &          \\
		\ \ \ \ \ \ \ Cor$(I_1,I_2)=-0.5$&          &          & -0.000257&          & -0.000255& -0.000253\\
		\ \ \ \ \ \ \ Cor$(I_1,I_2)=0$&          &          & -0.000237&          & -0.000259& -0.000177\\
		\ \ \ \ \ \ \ Cor$(I_1,I_2)=0.5$&          &          & -0.000259&          & -0.000254& -0.000264\\
		\ \ \ \ \ \ \ Clayton param. $2$        &          &   &-0.001198       &    & 0.000591 &-0.001815\\
		\ \ \ \ \ \ \ Clayton param. $4$        &          &   &-0.001300       &    & 0.001391 &-0.001737\\
		\ \ \ \ \ \ \ Clayton param. $6$        &          &   &-0.001216       &    & 0.001831 &-0.001516\\
		\cline{1-7}
		\textit{Normal model} &&&&&&\\
		&  0.016076&  0.016076&          &  0.016076&          &          \\
		\ \ \ \ \ \ \ Cor$(I_1,I_2)=-0.5$&          &          &  0.016071&          &  0.016079&  0.015937\\
		\ \ \ \ \ \ \ Cor$(I_1,I_2)=0$&          &          &  0.016069&          &  0.016082&  0.016054\\
		\ \ \ \ \ \ \ Cor$(I_1,I_2)=0.5$&          &          &  0.016085&          &  0.016050&  0.016098\\
		\ \ \ \ \ \ \ Clayton param. $2$        &          &   & 0.014804       &    & 0.020045 & 0.013340\\
		\ \ \ \ \ \ \ Clayton param. $4$        &          &   & 0.014809       &    & 0.022825 & 0.013694\\
		\ \ \ \ \ \ \ Clayton param. $6$        &          &   & 0.015078       &    & 0.025028 & 0.014333\\
		\cline{1-7}
		\textit{Log-normal model} &&&&&&\\
		&  0.042801&  0.042801&          &  0.042802&          &          \\
		\ \ \ \ \ \ \ Cor$(I_1,I_2)=-0.5$&          &          &  0.042768&          &  0.042844&  0.042255\\
		\ \ \ \ \ \ \ Cor$(I_1,I_2)=0$&          &          &  0.042780&          &  0.042849&  0.042723\\
		\ \ \ \ \ \ \ Cor$(I_1,I_2)=0.5$&          &          &  0.042800&          &  0.042781&  0.042798\\
		\ \ \ \ \ \ \ Clayton param. $2$        &          &   & 0.039364       &    & 0.053132 & 0.035512\\
		\ \ \ \ \ \ \ Clayton param. $4$        &          &   & 0.039402       &    & 0.060442 & 0.036422\\
		\ \ \ \ \ \ \ Clayton param. $6$        &          &   & 0.040109       &    & 0.066215 & 0.038124\\\bottomrule
	\end{tabular}
	\caption{\it \small Medians of $I^{cm}$, $I^c$ and $I$, and crossing points $d^{cm}$, $d^{c}$ and $d^{\star}$ of the pairs of cdf's $\left(F_{I},F_{I^{cm}}\right)$, $\left(F_{I},F_{I^{c}}\right)$ and $\left(F_{I^c},F_{I^{cm}}\right)$ for six different marginal distributions of $I_1$ and $I_2$, three different correlation coefficients, and three different parameters of the Clayton copula.}\label{Table1}
\end{table}

\subsection{Expected payoffs and dependence uncertainty spreads}\label{Sec54}

\begin{figure}[!h]
	\centering \includegraphics[width=1.02\textwidth]{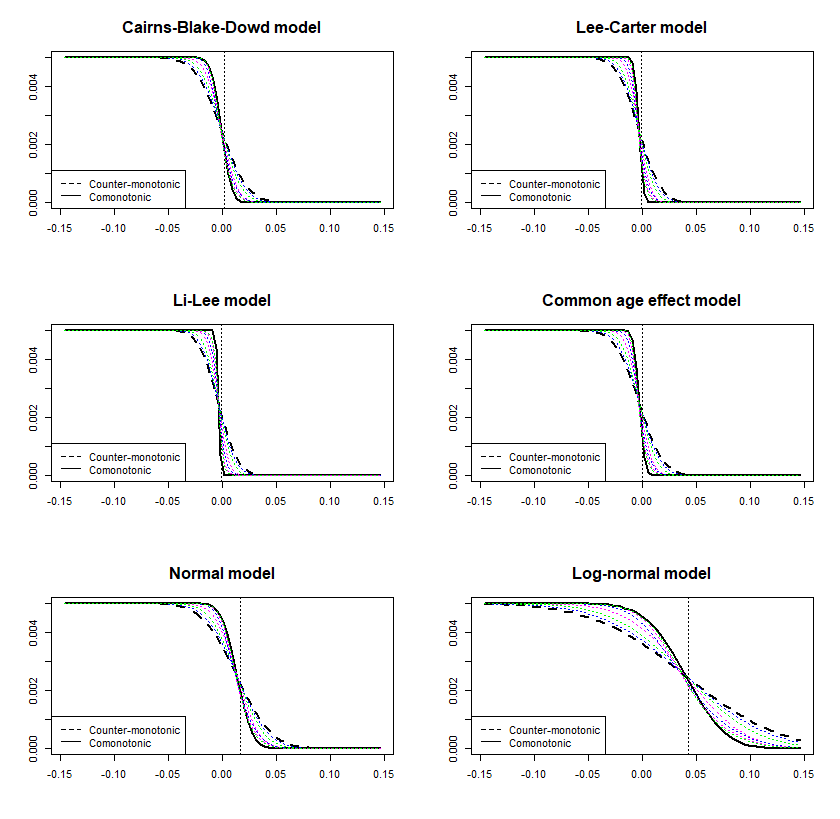}\caption{\it \small \textbf{Expected payoff of the longevity trend bond in function of the attachment point $\delta$} -- The exhaustion point is set to $\epsilon=\delta+0.005$. The thick straight curve corresponds to the expected payoff under comonotonicity, and the thick dashed curve corresponds to the expected payoff under countermonotonicity. The other curves correspond to the cases where $\left(I_1,I_2\right)$ has a Gaussian copula with correlation $-0.5$, $0$ or $0.5$, or a Clayton copula with parameter $2$, $4$ and $6$.}\label{Figure4}
\end{figure}
This subsection supplements the analysis by focusing on the expected payoffs and the dependence uncertainty spread. Figure \ref{Figure4} displays the expected payoff of the longevity trend bond in function of $\delta$, where $\epsilon-\delta=0.005$. The choice of $\epsilon-\delta$ is consistent with the original design of the Kortis bond with $\epsilon=3.9\%$ and $\delta=3.4\%$. The analysis is performed for $\delta \in[-15\%,15\%]$, which corresponds approximately to the range of values attained by the longevity divergence index $I^c$ across all models in this analysis. The thick straight curve corresponds to the expected payoff under comonotonicity, and the thick dashed curve corresponds to the expected payoff under countermonotonicity. The other curves correspond to the cases where $\left(I_1,I_2\right)$ has a Gaussian copula with correlation $-0.5$, $0$ or $0.5$, or a Clayton copula with parameter $2$, $4$ and $6$.

Recall from the results of Figures \ref{Figure1} and \ref{Figure2} that for all models, values of $\delta$ and $\epsilon$ above the median lead to $\mathbb{E}\left[r(I^c)\right]\leq \mathbb{E}\left[r(I)\right] \leq \mathbb{E}\left[r(I^{cm})\right]$, whereas values below the median lead to the opposite case, i.e. $\mathbb{E}\left[r(I^{cm})\right]\leq \mathbb{E}\left[r(I)\right] \leq \mathbb{E}\left[r(I^{c})\right]$. However, values near the median can lead to ambiguous situations, where the existence of only one, or none, of the bounds can be ascertained. The results on Figure \ref{Figure4} corroborate that conclusion. The ordering becomes clear further away from the median, but not around the median. Note that the neighborhood around the median where the ambiguity arises is influenced by the choice of $\epsilon$ and $\delta$.

In addition to illustrating the expected payoffs and the ordering relationships, Figure \ref{Figure4} also highlights that under all models, the dependence uncertainty spread is close to zero in two situations.

The first situation is when the bond is designed to hedge against `median' risk, i.e. values of $\delta$ near the median, with $\epsilon-\delta=0.005$. This situation is rather interesting because it is around the median where the ambiguity on the order of the expected payoffs arises. But because the cdf's cross near the median, the dependence uncertainty spread approaches zero in that neighborhood. Thus, the results in Figure \ref{Figure4} illustrate how, in spite of that ambiguity, the dependence model risk is small where the crossing occurs.

\begin{figure}[!h]
	\centering \includegraphics[width=1.02\textwidth]{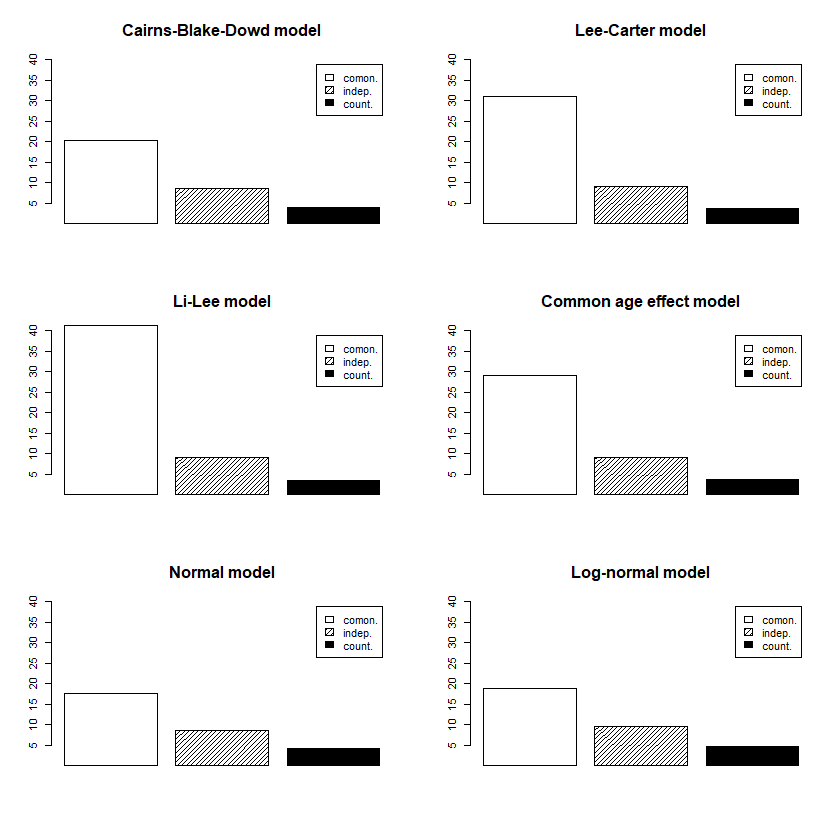}\caption{\it \small \textbf{Dependence uncertainty spread relative to the maximum payoff $\epsilon-\delta$ (in $\%$) for three values of $\delta$} -- Within each panel, the bars correspond to the ratio $100\times \frac{\mathbb{E}\left[r(I^{cm})\right]-\mathbb{E}\left[r(I^{c})\right]}{\epsilon-\delta}$. Each bar gives the value of that ratio for a given value of $\delta$, such that $\delta$ is equal to the 95th quantile of the longevity divergence index $I$ where $I_1$ and $I_2$ are either comonotonic (left white bar), independent (middle dashed bar), or countermonotonic (right black bar). The exhaustion point is set to $\epsilon=\delta+0.005$.}\label{Figure5}
\end{figure}

The second situation where all expected payoffs converge to the same value is when the bond is designed to hedge against tail risk, i.e. for low and large values of $\delta$ with $\epsilon-\delta=0.005$. Such designs are relevant as the issuer may be interested in hedging against large deviations between the mortality improvement indices of the two countries, as it was the case for Swiss Re's Kortis bond \citep{HuntBlake2015}. Figure \ref{Figure5} illustrates the dependence uncertainty spread for the right tail; similar results hold for the left tail. The figure provides the dependence uncertainty spread relative to the maximum payoff $\epsilon-\delta$ (in $\%$) for three values of $\delta$. Within each panel, the bars correspond to the ratio $100\times \frac{\mathbb{E}\left[r(I^{cm})\right]-\mathbb{E}\left[r(I^{c})\right]}{\epsilon-\delta}$. Each bar gives the value of that ratio for a given value of $\delta$, such that $\delta$ is equal to the 95th quantile of the longevity divergence index $I$ where $I_1$ and $I_2$ are either comonotonic (left white bar), independent (middle dashed bar), or countermonotonic (right black bar). Since the support of $I$ is larger when $I_1$ and $I_2$ are countermonotonic, the value of $\delta$ on the right bar is larger compared to the left bar. The figure thus shows how the dependence uncertainty spread decreases as the values of $\delta$ increase.

A caveat to this latter point is that the idea of hedging against tail basis risk itself is subject to the consequences of the uncertainty on the dependence between $I_1$ and $I_2$. In particular, since the support of $I$ becomes larger as the dependence becomes more negative, values of $\delta$ that allow to hedge against tail risk when $I_1$ and $I_2$ are countermonotonic (resp. comonotonic) may be too large (resp. low) if the true dependence structure of $(I_1,I_2)$ is comonotonic (resp. countermonotonic). This implies that, when the aim is to hedge against tail risk, there is a trade-off in the choice of $\delta$: high values allow to reduce the dependence uncertainty spread, but if they are too high, then the bond may not be relevant anymore because the layer that it intends to hedge could be too large for the support of $I$.

\subsection{Conclusion of the numerical study}\label{Sec55}
Overall, the analysis of the Kortis bond suggests that under the Gaussian copula, approximating the crossing points by the median of $I^{cm}$ is accurate for this data set and under the marginal distributions considered here. However, this approximation does not necessarily hold under non-symmetric copulas such as the Clayton copula. Nevertheless, the present analysis suggests that using extreme dependence structures as bounds in the context of dependence model risk is likely to be appropriate when the longevity trend bond is designed to hedge against tail risk, i.e.\ for sufficiently large or low values of $\delta$ and $\epsilon$. Further, the analysis of the expected payoffs suggests that in case the bond is designed to hedge against median risk, and to some extent against tail risk as well, the dependence uncertainty spread can be close to zero, meaning that dependence model risk can be low. However, for tail risk, the latter conclusion is subject to a trade-off. Another issue is that the values of the median depend on the choice of the marginal models, meaning that model risk on the marginal distributions can be substantial.

\section{Closing discussion}
\label{Sec-Conclusion}

This paper considers differences of stop-loss payoffs contingent on the difference of two random variables. Differences of two stop-loss payoffs are known in quantitative finance as bull call spreads, and in actuarial science as layer (re-)insurance or limited stop-loss. The recent life market has also witnessed an increase of securities with similar payoff structures. In particular, for longevity trend bonds, which gained popularity in the literature since the launch in 2010 of Swiss Re's Kortis bond, this payoff function involves a difference of two r.v.'s.

The aim is to study whether the situations where the two r.v.'s are perfectly dependent, i.e.\ comonotonic or countermonotonic, lead to upper and lower bounds for that expected payoff. In case they do, they can be relevant to determine bounds for the price, or to quantify the extent of dependence model risk using the dependence uncertainty spread. The analysis shows that (i) the design of the payoff function, and specifically the choice of the layers, determines the order, (ii) extreme dependence structures (i.e.\ comonotonicity and countermonotonicity) may bound from the same side, and (iii) these two properties are specific to the originally assumed dependence structure of the two r.v.'s. 

These findings raise the importance of studying the crossing points of the cdf of the original difference with the cdf's of its comonotonic and countermonotonic transforms. Some analytical results are derived, supplemented with a numerical study of the 2010 Swiss Re Kortis bond using different mortality models for the marginal distributions. The numerical study shows that the cdf's generally have unique pairwise crossing points. Under symmetric copulas, those crossing points can reasonably be approximated by the difference of the marginal medians, but this approximation is not necessarily valid for asymmetric copulas. Nevertheless, the analysis suggests that selecting the layers of the bond in the aim of hedging tail risk leads to a situation where the comonotonic and countermonotonic transforms can be appropriate bounds. In addition, the analysis of the expected payoffs suggests that the dependence uncertainty spread can be close to zero (i.e. low dependence model risk) when the layers are selected in the aim of hedging against median risk, and to some extent against tail risk as well.

An important remark from the numerical analysis is that although the general conclusions stated above are valid for all six marginal models, their outputs are different. The differences can be observed in the support of the longevity divergence index, and also in the numerical value of the median. The most striking result is for the log-normal model, for which the attachment and exhaustion points are below or near the crossing points, whereas for all other five models, the attachment and exhaustion points are above the crossing points. This means that the bounds are reversed, or even ambiguously ordered, for the log-normal models compared to the other five. These differences hint at a potential model risk regarding the marginal models. This issue is beyond the scope of the present paper, which focuses on model risk regarding the dependence model, but it is certainly a relevant topic for future research.

\bibliographystyle{agsm}
\bibliography{References_LTB}
\appendix
\renewcommand{\theequation}{\thesection.\arabic{equation}}

\section{Proof of Lemma \ref{Theorem}}\label{Theorem-proof}
The derivative of the function $f:x \mapsto \mathbb{E}\left[\left(Y-x\right)_+\right] - \mathbb{E}\left[\left(X-x\right)_+\right]$ is given by:
$$f^{\prime}(x) = F_{Y}(x) - F_{X}(x).$$
Since $X\preceq_{cx}Y$, the sets $D_{\leq}^{(j)}\left(X,Y\right)$ and  $D_{\geq}^{(j)}\left(X,Y\right)$ are not empty for $j=0,1,\ldots,\frac{N-1}{2}$. Further, by definition, $f$ is non-decreasing on each set $D_{\leq}^{(j)}\left(X,Y\right)$, and non-increasing on each set $D_{\geq}^{(j)}\left(X,Y\right)$. In particular, for a fixed $j\in \{0,1,\ldots,\frac{N-1}{2}\}$, if $\delta,\epsilon\in D_{\leq}^{(j)}\left(X,Y\right)$ with $\delta<\epsilon$, then $f$ is non-decreasing, and hence, $f(\delta)\leq f(\epsilon)$. Further, the inequality $f(\delta)\leq f(\epsilon)$ is equivalent to:
$$\mathbb{E}\left[\left(Y-\delta\right)_+\right] - \mathbb{E}\left[\left(X-\delta\right)_+\right] \leq \mathbb{E}\left[\left(Y-\epsilon\right)_+\right] - \mathbb{E}\left[\left(X-\epsilon\right)_+\right] \Longleftrightarrow \mathbb{E}\left[r(Y)\right] \leq \mathbb{E}\left[r(X)\right].$$
Therefore, if $\delta,\epsilon\in D_{\leq}^{(j)}\left(X,Y\right)$, then $\mathbb{E}\left[r(Y)\right] \leq \mathbb{E}\left[r(X)\right]$. Similarly, if $\delta,\epsilon \in D_{\geq}^{(j)}\left(X,Y\right)$ with $\delta<\epsilon$, $f$ is non-increasing, which leads to $\mathbb{E}\left[r(Y)\right] \geq \mathbb{E}\left[r(X)\right]$.

\section{Proof of Theorem \ref{Theorem_crossing}}\label{Theorem_crossing-proof}
Suppose first that $d_{YZ}=d_{XZ}=d$. It follows that $F_{X}(d)=F_{Z}(d)=F_{Y}(d)$, and since $F_{X}$ and $F_{Y}$ have a unique crossing point, then necessarily $d_{XY}=d_{XZ}=d_{YZ}=d$. Using the same argument for the cases where $d_{XY}=d_{YZ}$ and $d_{XY}=d_{XZ}$ proves that, in case of unique pairwise crossing points, two points coincide if and only if all three coincide.

Inequalities $d_{XY}<d_{XZ}<d_{YZ}$ and $d_{YZ}<d_{XZ}<d_{XY}$ are in general possible, as the inequalities of the cdf's before and after the crossing points resulting from $X \preceq_{cx} Y\preceq_{cx} Z$ do not conflict. 

The other inequalities are not possible. The proofs are all based on the same argument, which is detailed for one of them only. In particular, it is shown that the inequality $d_{XY}<d_{YZ}<d_{XZ}$ cannot hold, and similar reasoning applies to the three other inequalities $d_{XZ}<d_{YZ}<d_{XY}$, and $d_{XZ}<d_{XY}<d_{YZ}$ and $d_{YZ}<d_{XY}<d_{XZ}$.

Suppose that $d_{XY}< d_{YZ}< d_{XZ}$. Since $X \preceq_{cx} Y$ and $X \preceq_{cx} Z$, and the three cdf's $F_X$, $F_Y$ and $F_Z$ have unique pairwise crossing points, then:
\begin{equation}\label{Ap5-1}F_Y(x) \leq F_{X}(x) \leq F_{Z}(x), \quad \text{for all } d_{XY} \leq x \leq d_{XZ}.\end{equation}
However, for $d_{YZ}\leq x \leq d_{XZ}$, the inequality $Y\preceq_{cx}Z$ leads to:
$$F_{Z}(x) \leq F_{Y}(x),$$
which, taking \eqref{Ap5-1} into account, is possible only for $x=d_{XY}=d_{XZ}=d_{YZ}$. Thus, $d_{XY}< d_{YZ}< d_{XZ}$ is not possible.

\section{Proof of Theorem \ref{Theorem_Disp}}\label{Theorem_Disp-proof}
Let $d^{\star}$ be a crossing point of $F_{I^{c}}$ and $F_{I^{cm}}$, and define $p^{\star}$ such that:
$$p^{\star} = F_{I^{c}}\left(d^{\star}\right) = F_{I^{cm}}\left(d^{\star}\right).$$
Since the cdf's are strictly increasing, then:
\begin{equation}\label{C1}
d^{\star} = F^{-1}_{I^{c}}\left(p^{\star}\right) =F^{-1}_{I^{cm}}\left(p^{\star}\right).\end{equation}
Consider the function $f:p \mapsto F^{-1}_{I_1}(p) - F^{-1}_{I_2}(1-p)$, with $I^{cm}\overset{d}{=}f(U)$ where $U$ is uniformly distributed over $[0,1]$. The function $f$ is always non-decreasing, and hence, from \cite{Dhaene2002a}, the following equality holds:
\begin{equation}\label{C2}
	F^{-1}_{I^{cm}}\left(p^{\star}\right)=F^{-1}_{I_1}\left(p^{\star}\right) - F^{-1}_{I_2}\left(1-p^{\star}\right).\end{equation}
Consider the function $g:p\mapsto F^{-1}_{I_1}(p) - F^{-1}_{I_2}(p)$, where $I^{c}\overset{d}{=}g(U)$. Suppose that $I_2 \preceq_{DISP} I_1$, then by definition, the function $g$ is non-decreasing, and hence:
\begin{equation}\label{C3}
	F^{-1}_{I^{c}}\left(p^{\star}\right)=F^{-1}_{I_1}\left(p^{\star}\right) - F^{-1}_{I_2}\left(p^{\star}\right).\end{equation}
Plugging \eqref{C2} and \eqref{C3} in \eqref{C1} leads to:
$$
	F^{-1}_{I_1}\left(p^{\star}\right) - F^{-1}_{I_2}\left(1-p^{\star}\right)=F^{-1}_{I_1}\left(p^{\star}\right) - F^{-1}_{I_2}\left(p^{\star}\right) \Longleftrightarrow F^{-1}_{I_2}\left(1-p^{\star}\right)= F^{-1}_{I_2}\left(p^{\star}\right),$$
and the strict increasingness of $F^{-1}_{I_2}$ proves that there is a single crossing point $d^{\star}$ between $F_{I^c}$ and $F_{I^{cm}}$ such that $F_{I^{c}}\left(d^{\star}\right) = F_{I^{cm}}\left(d^{\star}\right)=\frac{1}{2}$. In particular, the medians of $I^c$ and $I^{cm}$ are equal to $d^{\star}$. 

Suppose now that $I_1 \preceq_{DISP} I_2$, which means that $g$ is non-increasing, i.e.\ $F_{I^c}^{-1}(p)=g\left(1-p\right)$ for $p\in(0,1)$. This equality is equivalent to 	$F^{-1}_{I^{c}}\left(p^{\star}\right)=F^{-1}_{I_1}\left(1-p^{\star}\right) - F^{-1}_{I_2}\left(1-p^{\star}\right)$, and combined with \eqref{C1} and \eqref{C2}, it follows that:
$$F^{-1}_{I_1}\left(p^{\star}\right)=F^{-1}_{I_1}\left(1-p^{\star}\right),$$
i.e.\ there is again a single crossing point between $F_{I^c}$ and $F_{I^{cm}}$ given by $d^{\star}$, such that $p^{\star}=\frac{1}{2}$.

\section{Proof of Theorem \ref{Theorem_Sym}}\label{Theorem_Sym-proof}
The proof is split into two parts. The first part proves that $F_{I^{c}}$ and $F_{I^{cm}}$ have a unique crossing point which is equal to $\bar{\mu}$. The second part proves that the pair of cdf's $\left(F_{I},F_{I^{cm}}\right)$ and $\left(F_{I},F_{I^{c}}\right)$ each have a unique crossing point, and that these two points are equal. Note that Theorem \ref{Theorem_crossing} guarantees that in case the crossing points are unique and that two points coincide, then all three pairwise crossing points are equal. Nevertheless, it is necessary to prove that all three pairwise crossing points are unique.

\subsection{Crossing points of $F_{I^c}$ and $F_{I^{cm}}$}
Let $\bar{\mu} = \mu_1-\mu_2$, and let $I\overset{d}{=}\bar{\mu} + \sigma_1W_1 - \sigma_2W_2$, where $W_1$ and $W_2$ are identically distributed with cdf $F_W$. Consider the function $g:u\mapsto F^{-1}_{I_1}(u)-F^{-1}_{I_2}(u)=\bar{\mu} + \left(\sigma_1-\sigma_2\right)F_W^{-1}(u)$. The function $g$ is always monotone, and hence, $I_1$ and $I_2$ are ordered in the dispersive sense. From Theorem \ref{Theorem_Disp}, there is a unique crossing point $d^{\star}$ between $F_{I^c}$ and $F_{I^{cm}}$ that corresponds to their median. Further, since $I^c \overset{d}{=}g(U)$, and $F_W$ is centered and symmetric about $0$, then $d^{\star}=\bar{\mu}$.

\subsection{Crossing points of $\left(F_{I^c},F_{I}\right)$ and of $\left(F_{I^{cm}},F_{I}\right)$}
This part of the proof shows first that $\bar{\mu}$ is a crossing point for all three cdf's, and second that it is a unique crossing point.

Consider the difference $Y\overset{d}{=}\tilde{W}_1-\tilde{W}_2$, with $\tilde{W}_i\overset{d}{=}\sigma_iW_i$, where the random vector has a radially symmetric copula $C$ with density $c$. The density of the difference $Y$ is given by:
$$f_{Y}(y) = \int_{-\infty}^{+\infty} f_{\left(\tilde{W}_1,\tilde{W}_2\right)}(x,x-y)\text{d}x,$$
where $f_{\left(\tilde{W}_1,\tilde{W}_2\right)}$ is the bivariate density function of the vector $\left(\tilde{W}_1,\tilde{W}_2\right)$. Since copula functions are invariant by strictly increasing transformation, then:
$$\mathbb{P}\left[\tilde{W}_1\leq w_1,\tilde{W}_2\leq w_2\right]=C\left(F_{\tilde{W}_1}\left(w_1\right),F_{\tilde{W}_2}\left(w_2\right)\right),$$
where $C$ is the copula of $\left(W_1,W_2\right)$, and hence:
$$f_{Y}(y) = \int_{-\infty}^{+\infty} f_{\tilde{W}_1}(x)f_{\tilde{W}_2}(x-y)c\left(F_{\tilde{W}_1}\left(x\right),F_{\tilde{W}_2}\left(x-y\right)\right)\text{d}x.$$
The symmetry of $W_1$ and $W_2$ about $0$ means that $f_{\tilde{W}_i}(x)=f_{\tilde{W}_i}(-x)$, as well as $F_{\tilde{W}_i}(x) = 1 - F_{\tilde{W}_i}(-x)$, whereas the radial symmetry of $C$ means that $c(u,v)=c(1-u,1-v)$. Therefore:
$$f_{Y}(y) = \int_{-\infty}^{+\infty} f_{\tilde{W}_1}(-x)f_{\tilde{W}_2}(-x+y)c\left(F_{\tilde{W}_1}\left(-x\right),F_{\tilde{W}_2}\left(-x+y\right)\right)\text{d}x,$$
and a change of variable from $-x$ to $x$ leads to:
$$f_{Y}(y) = \int_{-\infty}^{+\infty} f_{\tilde{W}_1}(x)f_{\tilde{W}_2}(x-(-y))c\left(F_{\tilde{W}_1}\left(x\right),F_{\tilde{W}_2}\left(x-(-y)\right)\right)\text{d}x,$$
which means that $f_{Y}(y)=f_{Y}(-y)$, and hence, that $Y$ is symmetric about $0$. In particular, the difference $I=I_1-I_2$ can be written as:
$$I\overset{d}{=}\bar{\mu} + Y,$$
where $Y$ is symmetric about it's mean $0$ and has a standard deviation $\sigma_{Y}>0$. Therefore, the crossing points of $F_I$ and $F_{I^{cm}}$ satisfies:
\begin{equation}\label{E1}F_{Y}(x-\bar{\mu}) = F_{W}\left(\frac{x-\bar{\mu}}{\sigma_1+\sigma_2}\right).\end{equation}
For $\sigma_1\neq \sigma_2$, the crossing points of $F_I$ and $F_{I^{c}}$ satisfies:
\begin{equation}F_{Y}(x-\bar{\mu}) = F_{W}\left(\frac{x-\bar{\mu}}{\sigma_1-\sigma_2}\right).\end{equation}
It follows that $\bar{\mu}$ is a crossing point of all three cdf's. The remaining task is to prove that this crossing point is unique for $\sigma_1\neq \sigma_2$. Note that if $\sigma_1=\sigma_2$, $I^c$ is a degenerate r.v.\ whose cdf jumps at $\bar{\mu}$. This means that the crossing point of $F_{I^c}$ and $F_{I}$ is $\bar{\mu}$, and it is unique.

Suppose that $F_I$ and $F_{I^{cm}}$ have two additional crossing points $d_{1}^{cm}$ and $d_{2}^{cm}$. Suppose also that $d_2^{cm}>\bar{\mu}$, and let $\zeta = d_2^{cm}-\bar{\mu}$. It follows from \eqref{E1} that $F_I\left(d_2^{cm}\right)=F_{I^{cm}}\left(d_2^{cm}\right)$ is equivalent to $F_Y\left(\zeta\right) = F_W\left(\frac{\zeta}{\sigma_1+\sigma_2}\right)$,
and by symmetry, the equality $F_Y\left(-\zeta\right) = F_W\left(-\frac{\zeta}{\sigma_1+\sigma_2}\right)$ also holds. Thus, $d_1^{cm}=\bar{\mu} - \zeta$, which means that all additional pairs of crossing points are equidistant from $\bar{\mu}$. A similar reasoning applies to $F_I$ and $F_{I^c}$ if they have three or more crossing points.

Let $\bar{\mu} \pm \zeta $ be the two additional crossing points of $F_{I}$ and $F_{I^{cm}}$ on top of $\bar{\mu}$, and let $\bar{\mu} \pm \lambda$ be the two additional crossing points of $F_{I}$ and $F_{I^{c}}$ on top of $\bar{\mu}$. For any $x$ such that $ \bar{\mu} - \min\{\zeta,\lambda\} \leq x \leq \bar{\mu}$, it holds that $F_{I^c}(x)\leq F_{I^{cm}}(x)$ because $I^c \preceq_{cx}I^{cm}$, as well as $F_I(x)\leq F_{I^c}(x)$ because $I^c \preceq_{cx} I$, and $F_{I^{cm}}(x)\leq F_I(x)$ because $I\preceq_{cx}I^{cm}$. Combining these inequalities leads to:
$$F_{I^{cm}}(x)\leq F_I(x)\leq F_{I^c}(x) \leq F_{I^{cm}}(x), \quad \text{for all } x\in[\bar{\mu} - \min\{\zeta,\lambda\},\bar{\mu}].$$
This inequality holds only if $F_I(x)= F_{I^c}(x)= F_{I^{cm}}(x),$ for all $x\in[\bar{\mu} - \min\{\zeta,\lambda\},\bar{\mu}],$ and since the cdf's are strictly increasing, it necessarily follows that $\min\{\zeta,\lambda\}=0$. The case where $x\in[\bar{\mu},\bar{\mu}+ \max\{\zeta,\lambda\}]$ leads to the conclusion that $\max\{\zeta,\lambda\}=0$. Therefore, it follows that $F_I$ and $F_{I^{cm}}$, as well as $F_I$ and $F_{I^{c}}$, have a single crossing point $\bar{\mu}$, and hence, that all three cdf's cross once at the same point which is the median. This ends the proof.

\section{Marginal models}\label{AppendixModels}
\textbf{Cairns-Blake-Dowd model}

\cite{CBD2006}'s two-factor stochastic mortality model is based on the observation that at any time, the cross-sectional mortality odd ratio can be approximated by a linear function of age. Under the Cairns-Blake-Dowd (CBD) model, $\mu_{x,t}^{(i)} = \log \left(1 + \exp\left(\theta_t^{(i)} + x\tau_t^{(i)}\right)\right).$
For each $i$, the time-dependent parameters $\theta_t^{(i)}$ and $\tau_t^{(i)}$ are modeled using correlated random walks with drift, which is a standard specification  \citep{PitaccoDenuit,Enchev2017}. \\

\textbf{Lee-Carter model}

Under the Lee-Carter model, the central death rate $\mu_{x,t}^{(i)}$ at age $x$ and time $t$ in population $i$ is defined as $\mu_{x,t}^{(i)} = \exp\left(\lambda_{x}^{(i)} + \beta_{x}^{(i)}\kappa_{t}^{(i)}\right).$ For each age $x$, $\lambda_{x}^{(i)}$ represents the age-dependent average log-rates. For each time $t$, $\kappa_{t}^{(i)}$ captures the mortality improvement in population $i$. For each age $x$, $\beta_{x}^{(i)}$ captures the sensitivity of that age group to the time component. The time-dependent parameters $\kappa_t^{(i)}$ are assumed to follow random walks with drift.\\

\textbf{Li-Lee model}

The Li-Lee model extends the Lee-Carter model by adding a common time effect, such that $\mu_{x,t}^{(i)}$ is given by $\mu_{x,t}^{(i)} = \exp\left(\lambda_{x}^{(i)} + \beta_{x}^{(i)}\kappa_{t}^{(i)} + \bar{\beta}_{x}\bar{\kappa}_t\right)$. The component $\bar{\beta}_{x}\bar{\kappa}_t$ measures the mortality improvement which is common to both populations for a given age $x$, whereas $\beta_{x}^{(i)}\kappa_{t}^{(i)}$ captures the mortality improvement that is specific to population $i$ and age $x$. Random walks with drift are used to model $\kappa_{t}^{(1)}$, $\kappa_{t}^{(2)}$ and $\bar{\kappa}_{t}$.\\

\textbf{Common age effect model}

\cite{Kleinow2015}'s common age effect (CAE) model is such that $\mu_{x,t}^{(i)} = \exp\left(\lambda_{x}^{(i)} + \beta_{x}\kappa_{t}^{(i)} + \bar{\beta}_{x}\bar{\kappa}^{(i)}_t\right)$. The difference with the Li-Lee model is that the age-dependent parameters $\beta_x$ and $\bar{\beta}_x$ are common to both populations, and the time-dependent parameters $\kappa_{t}^{(i)}$ and $\bar{\kappa}_{t}^{(i)}$ are specific to each population. Again, the time-dependent parameters are modeled using ARIMA(0,1,0) with drift.\\

\textbf{Normal and log-normal models}

Whereas the four models above focus on the central death rates by age, the last two models are fitted directly to the following time series:
\begin{equation}\label{TS}
	y_t^{(i)} = 1 - \frac{1}{\omega^{(i)}-\alpha^{(i)} + 1}\underset{x=\alpha^{(i)}}{\overset{\omega^{(i)}}{\sum}} \left(\frac{\mu_{x,t+T}^{(i)}}{\mu_{x,t}^{(i)}}\right)^{\frac{1}{T}}.
\end{equation}
Under the fifth model, $y_t^{(i)}$ is assumed to follow an auto-regressive process, and hence, the indices $I_1$ and $I_2$ follow a normal distribution. Under the sixth model, $\log\left(y^{(i)}_t\right)$ is assumed to follow an ARIMA(0,1,0), and hence, the indices $I_1$ and $I_2$ follow a log-normal distribution.\\

\textbf{Data and computations}

The estimation uses the total number of deaths and exposure-to-risk from the Human Mortality Database (\url{www.mortality.org}) of England and Wales and the US. The ages selected in the fitting process are 25-95. The observation period is 1950-2009. The observations are limited to 2009 in order to provide an analysis which is consistent with the issue date of the Kortis bond. 

For the CBD model, the parameters $\theta_t^{(i)}$ and $\tau_t^{(i)}$ are estimated using repeated OLS from the cross-section of mortality odd ratios at different times. More details on the CBD model can be found in \cite{PitaccoDenuit}. The parameters $\beta$, $\bar{\beta}$, $\kappa$ and $\bar{\kappa}$ in the Lee-Carter, the Li-Lee, and the CAE are estimated using maximum-likelihood with Poisson distribution for the number of deaths, and are subject to the standard identifiability constraints; see e.g.\ \cite{PitaccoDenuit} for more details on estimating the Lee-Carter model, and \cite{Enchev2017} for the Li-Lee and the CAE models.

All calculations are performed using \texttt{R} software \citep{R}, and the packages \texttt{shiny} and \texttt{rsconnect} were used for the design and deployment of the online tools \citep{shiny,rsconnect}.
\end{document}